\documentclass[prl,showpacs,preprintnumbers,amsmath,amssymb,twocolumn,superscriptaddress,notitlepage,nofootinbib]{revtex4-2}
\usepackage{graphicx,color}
\usepackage{graphicx}
\usepackage{amsmath,amssymb,amsfonts}
\usepackage{stackrel}
\usepackage{enumitem}
\usepackage[english]{babel}
\usepackage{verbatim}
\usepackage{comment}
\usepackage{xcolor}
\usepackage{subcaption}
\usepackage{dcolumn}
\usepackage{physics}
\usepackage{verbatim}
\usepackage{float}
\usepackage{color}
\usepackage{multirow}
\usepackage{graphicx}
\usepackage{epstopdf}
\usepackage{epsfig}
\usepackage{slashed}
\usepackage{physics}
\usepackage{multirow}
\usepackage{makecell}
\usepackage{placeins}
\usepackage{setspace}
\usepackage{tabularx}
\usepackage{colortbl}
\usepackage{float}
\usepackage{placeins}
\usepackage{ragged2e}
\usepackage{bm}

\newcommand{\beq}{\begin{equation}}
\newcommand{\eeq}{\end{equation}}
\newcommand{\bal}{\begin{align}}
\newcommand{\eal}{\end{align}}
\newcommand{\bit}{\begin{itemize}}
\newcommand{\eit}{\end{itemize}}
\newcommand{\ben}{\begin{enumerate}}
\newcommand{\een}{\end{enumerate}}

\renewcommand{\eqref}[1]{Eq.~(\ref{eq:#1})}

\newcommand{\figref}[1]{Fig.~\ref{fig:#1}}

\newcommand{\tabref}[1]{Tab.~\ref{tab:#1}}

\definecolor{darkred}{rgb}{.7,.1,.1}
\usepackage{hyperref}
\hypersetup{colorlinks=true,
	breaklinks=true,
	pdfstartview=Fit,
	linkcolor=blue,
	citecolor=blue,
	urlcolor=blue}
\usepackage{cleveref}

\graphicspath{
  {./}
  {figures/}
}

\definecolor{dark-green}{rgb}{0.1,0.7,0.3}

\begin{document}

\title{Probing Collapsed Dark Matter Halos with Fast Radio Bursts}

\author{Yuxuan He}
\email{he.yx25@cityu.edu.hk}
\affiliation{Department of Physics, City University of Hong Kong, Kowloon, Hong Kong SAR, China}
\author{Weiyang Wang}
\email{wywang\_astroph@pku.edu.cn}
\affiliation{University of Chinese Academy of Sciences, Beijing 100049, China}
\author{Chen Zhang}
\email{zhangvchen@tongji.edu.cn}
\affiliation{School of Physics Science and Engineering, Tongji University, Shanghai 200092, China}
\affiliation{The HKUST Jockey Club Institute for Advanced Study, The Hong Kong University of Science and Technology, Hong Kong SAR, China}
\author{Yi-Ming Zhong}
\email{yiming.zhong@cityu.edu.hk}
\affiliation{Department of Physics, City University of Hong Kong, Kowloon, Hong Kong SAR, China}

\date{\today}

\begin{abstract}
Observations of ultra-dense substructures in strong lensing systems challenge the standard cosmological model at small scales.  Self-interacting dark matter (SIDM), as an alternative to the cold and collisionless dark matter (CDM) of the standard cosmological model, provides a natural mechanism for forming such structures via gravothermal core collapse. We show that strong gravitational lensing of fast radio bursts (FRBs) provides an effective approach to detecting these substructures and probing dark matter self-interactions. Core-collapsed SIDM halos exhibit steeper central density profiles than CDM halos, enhancing the lensing cross section and producing longer time delays between FRB images. We compute lensing properties of core-collapsed subhalos and host halos, including maximal impact parameters and time-delay distributions. We demonstrate that future all-sky monitors, such as BURSTT, SKA2-Low, and SKA2-Mid, which are expected to detect $10^{5}$--$10^{7}$ FRBs over a decade, can measure time-delay distributions with high statistical significance. Modeling collapsed halos with a cored power-law density profile with inner slope $\gamma=3$ and assuming no excess beyond the singular isothermal sphere lens model, we show that our strategy can probe self-interaction cross section strengths of $\sigma_{\text{SI}}/m \gtrsim \min\{18,\, 40\lambda_{\text{sub}}\}\,\text{cm}^2/\text{g}$, where $\lambda_{\text{sub}}$ parameterizes the collapse time of a subhalo relative to that of the isolated case.
\end{abstract}

\pacs{}
\maketitle

%%%%%%%%%%%%%%%%%%%%%%%%%%%%%%%%%%%%%%%%%%%%%%%%%%%%%%%%%%%%%%%
\noindent \textbf{\textit{Introduction}}. The standard cosmological model, $\Lambda$CDM, successfully describes the universe on large scales with wavenumber $k\lesssim \mathcal O (1)\,\text{Mpc}^{-1}$ (see, e.g.,~\cite{Planck:2018vyg}). However, it faces challenges on small scales (see, e.g., reviews~\cite{Bullock:2017xww, Tulin:2017ara}). In addition, strong lensing observations reveal ultra-dense substructures~\cite{Dutra:2024qac,Powell:2025rmj,Tajalli:2025qjx,Enzi2025,He:2025wco,Minor:2020hic,Vegetti2010,Vegetti:2026mmx} that are difficult to reconcile with  the cold and collisionless dark matter (CDM) of $\Lambda$CDM, which predicts halos with universal Navarro-Frenk-White density profiles~\cite{Navarro:1995iw, Navarro:1996gj}. Self-interacting dark matter (SIDM) has  emerged as a compelling alternative (see review~\cite{Tulin:2017ara}). The self-interactions of dark matter drive gravothermal evolution, leading first to core formation and later to core collapse of the halo~\cite{Essig:2018pzq,Gurian:2025zpc,Fischer:2025rky,Zeng:2023fnj,Yang:2023jwn}. While core formation can address the core-cusp and too-big-to-fail problems~\cite{deBlok:2009sp,Boylan-Kolchin:2011qkt}, core collapse produces halos with ultra-dense central regions that yield distinctive observational signatures. In particular, they can be probed through gravitational lensing: their compact centers enhance strong lensing of quasars~\cite{Gilman:2021sdr,Gilman:2022ida,Yu:2025tmp,Jiang:2025jtr,Kong:2025sqx,Hou:2025gmv}, while their outer profiles can be tested with weak lensing~\cite{DES:2021qzb,Adhikari:2024aff}.

\begin{figure}
    \centering
    \includegraphics[width=1.0\linewidth]{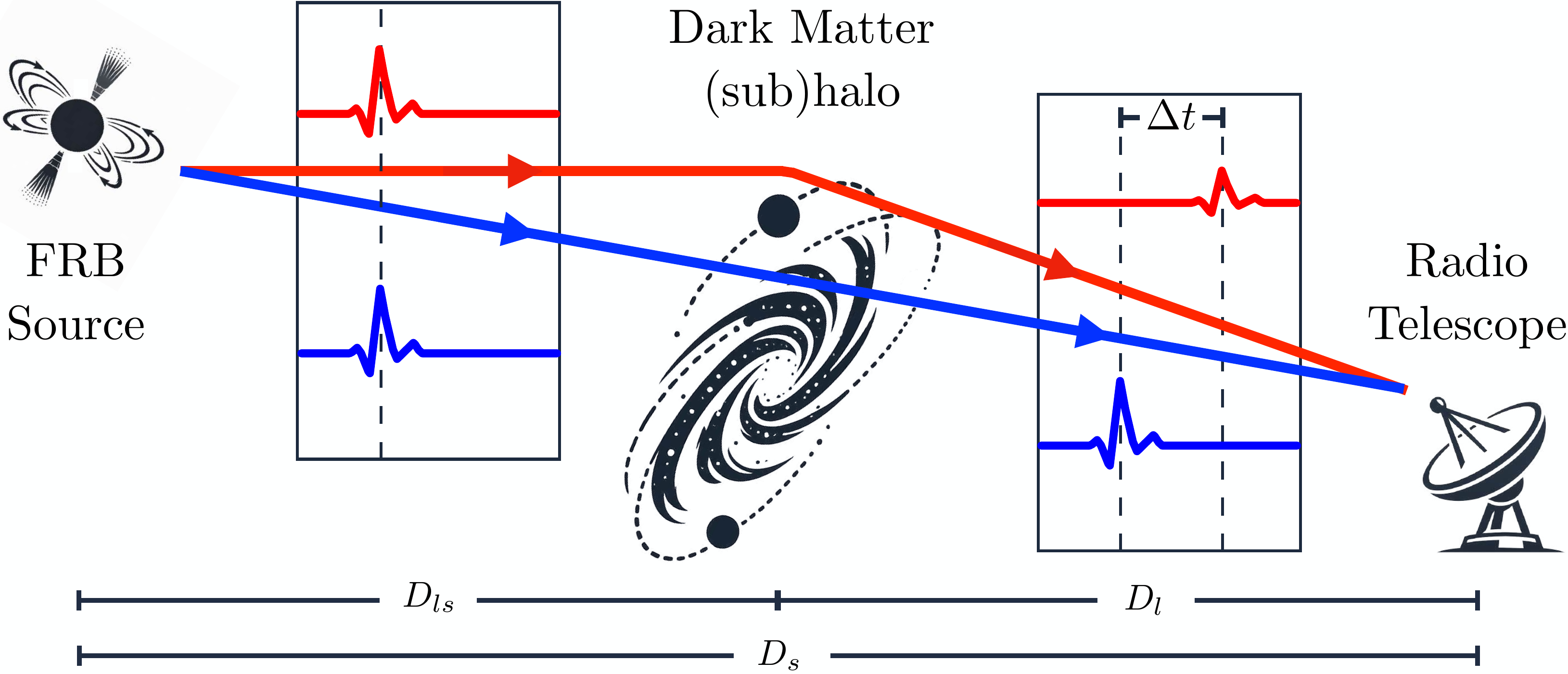}
    \caption{A schematic plot of the lensing of an FRB by core-collapsed self-interacting dark matter (sub)halos. The lensed FRB signal (red) is delayed with respect to the unlensed one (blue), resulting in multiple FRB images.}
    \label{fig:sch}
\end{figure}

Here we focus on a new observable---gravitational lensing of fast radio bursts (FRBs). FRBs are luminous, millisecond-duration radio transients with exceptionally high brightness temperatures~\cite{Zhang:2022uzl,Petroff:2019tty}. Their cosmological origin was definitively established through the identification of their host galaxies~\cite{2017Natur.541...58C,2019Sci...365..565B,2024ApJ...967...29L}. Although their physical origin remains uncertain~\cite{2023RvMP...95c5005Z}, FRBs have 
been demonstrated to be powerful probes of the universe \cite{Munoz:2016tmg,Li:2017mek,Wang:2018ydd,Wucknitz:2020spz,Adi:2021uuw, Leung:2022vcx,Singh:2023hbd,DallOsso:2024sjv,Pastor-Marazuela:2024lcd,Kader:2024uqm,Xiao:2024qay,Xiong:2025gtw,Meena:2025cdo,Li:2025wdc}  and may offer new insight on ultra-dense substructures. Similarly to lensing of quasars and galaxies, the ultra-dense cores of collapsed halos can lens FRBs, producing a longer time delay between the  FRB lensing images than in the CDM case. We focus on the regime in which individual halos dominate the delay, as shown in~\figref{sch}. To estimate the detection sensitivity, we adopt a cored power-law profile for collapsed halos. We first compute their lensing properties and the distribution of image time delays, and then perform parameter inference for a phenomenological two-parameter description of SIDM halo core collapse for future FRB observatories.

\noindent \textbf{\textit{Lensing Properties of core-collapsed SIDM halos}}. CDM halos are described by the NFW profile,
\begin{equation}
    \rho=\rho_s (r/r_s)^{-1}(1+r/r_s)^{-2},
    \label{eq:nfw}
\end{equation}
where $\rho_s$ is the scale density and $r_s\equiv r_{200}/c_{200}$ is the scale radius. Here $r_{200}$ is the radius within which the mean density is 200 times the critical density, and $c_{200}$ is the concentration, which we determine using the mass-concentration relation of Ref.~\cite{Ludlow:2016ifl}. The inner density of the profile~(\ref{eq:nfw}) scales as $\rho\propto r^{-1}$. By contrast, simulations~\cite{Zeng:2021ldo,Turner:2020vlf,Kahlhoefer:2019oyt,Jiang:2022aqw,Essig:2018pzq} show that core-collapsed SIDM halos develop a much cuspier profile, which can be modeled by the cored power law~\cite{Gilman:2022ida},
\begin{equation}
    \rho(r)=\rho_0 \left(1+{r^2}/{r_c^2}\right)^{-\gamma/2},
\label{eq:prof}
\end{equation}
where $\gamma$ characterizes the inner slope (a.k.a., cuspiness) and is typically in the range $2\lesssim\gamma\lesssim3$~\cite{Essig:2018pzq,Kahlhoefer:2019oyt,Turner:2020vlf,Zeng:2021ldo,Jiang:2022aqw,Yang:2022mxl}. $r_c$ is the collapsed core radius, which corresponds to the size of the secondary core of a collapsed halo. We adopt $r_c/r_s \simeq 10^{-2}$ from the fluid simulations~\cite{Pollack:2014rja, Essig:2018pzq, Outmezguine:2022bhq, Gilman:2022ida}. $\rho_0$ is the central density, fixed by requiring that the enclosed mass of profile~(\ref{eq:prof}) within $r\simeq 2r_s$ matches that of profile~(\ref{eq:nfw}) within the same radius~\cite{Yang:2022mxl, Yang:2023stn}.

\begin{figure}
    \centering       \includegraphics[width=0.96\linewidth]{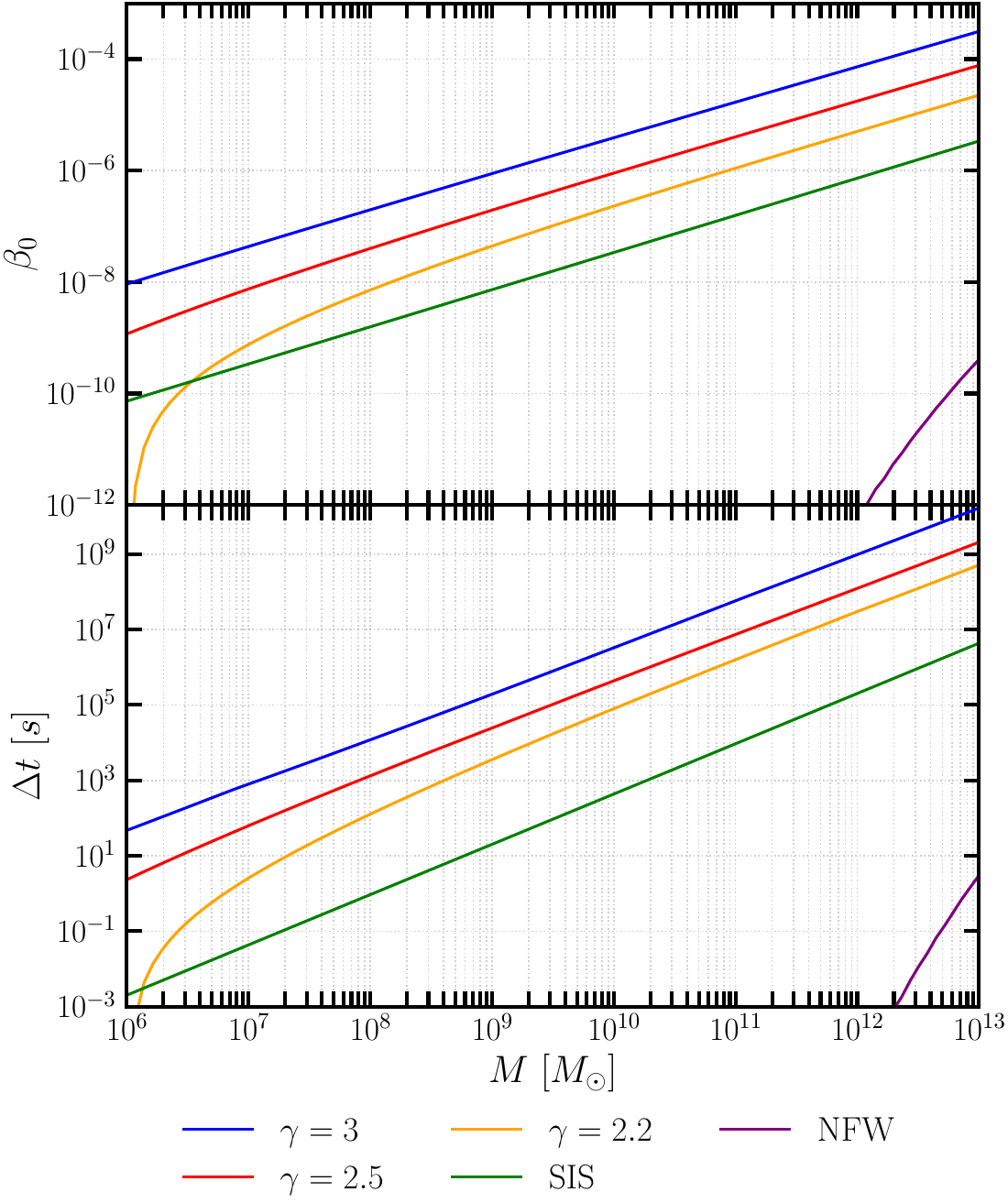}
    \caption{ (\emph{Upper}) The maximal impact angles $\beta_0$ that enable strong lensing to happen for various halo profiles and masses. (\emph{Lower}) The corresponding typical time delay $\Delta t$ for an inclination angle of e.g., $\beta/\beta_0=2/3$. }
    \label{fig:prof}
\end{figure}

For an individual halo with profile  (\ref{eq:prof}), strong lensing produces multiple images when the impact angle $\beta$ of an FRB source is smaller than the maximum impact angle $\beta_0$. As shown in the upper panel of Fig.~\ref{fig:prof} (with source redshift $z_s =1$ and lens redshift $z_l =0.75$), $\beta_0$ increases with both $\gamma$ and the lens mass $M$. For $2\leq \gamma\leq 3$, the strong lensing probability for profile~(\ref{eq:prof}) is significantly larger than that of profile~(\ref{eq:nfw}) at all masses. For our lens model~\eqref{prof}, three FRB images form when $\beta < \beta_0$ and we focus on the two brightest, corresponding to the minimum and the saddle point of the Fermat potential~\cite{Schneider:1992bmb}. The lower panel of Fig.~\ref{fig:prof} shows that their time delays $\Delta t$ also increase with $\gamma$ and $M$, so a cuspier halo produces longer time delays.

These individual halo results can be straightforwardly extended to a population of halos along the FRB propagation path, assuming that strong lensing occurs at most once and each halo lenses independently. We distinguish two types of lenses. The first is the host halos, with typical masses $M_{\text{host}} > 10^{10} M_{\odot}$. We model them as singular isothermal spheres (SIS), since these halos usually host galaxies and behave as SIS in observed strong lensing systems. The second is the subhalos, with typical masses in the range $10^6M_{\odot}<M_{\text{sub}}<10^{10}M_{\odot}$. We model them by either the NFW profile (\ref{eq:nfw}) for non-collapsed halos or the profile (\ref{eq:prof}) for collapsed halos.\footnote{Non-collapsed SIDM halos can also be in the core-formation stage, exhibiting a profile with a large core $r_c \approx 2 r_s$. Modeling them with the NFW profile leads to a negligible increase in the lensing probability, as indicated by~\figref{prof}.}

\noindent \textbf{\textit{FRB observations}}. \label{sec:obs}
We now describe the FRB source distribution and the detection of lensed FRBs at observatories.

\textit{(1) The redshift and luminosity distributions of FRB sources.}
The redshift and luminosity distributions of FRBs have been widely discussed in the literature~\cite{Shin:2022crt,Gupta:2025jyw,Lei:2025wyw,Chen:2024fne}, largely through data-driven approaches. Following Ref.~\cite{Luo:2020wfx}, we assume that the cumulative redshift-luminosity distribution factorizes as $\Phi( L_s, z) = \psi(L_s) \phi(z)$,\footnote{This assumption allows one to apply the Eddington $C^-$ method for completeness correction.} where $\psi(L_s) \propto L_s^{\alpha_L}$ is the cumulative luminosity distribution with $\alpha_L=-0.79$ and the source luminosity $L_s$ in the range of $ 2 \times 10^{39} < L_s < 3 \times 10^{44} \ \text{erg/s}$,\footnote{The lower cutoff on $L_s$ is set to correctly reproduce SKA2 event rates, as SKA2 has a lower luminosity threshold.} and $\phi(z)$ is the cumulative  redshift distribution, whose probability density is \emph{uniform}  in comoving volume $V$, i.e.,
${d \phi(z)}/{ d z} \propto {d V}/((1+z)dz)$.

\textit{(2)  Detection of lensed FRBs at observatories.}
 We focus on observatories with larger fields of view (FOVs) in order to maximize the number of lensing events. When the time delays between lensed images exceed the scan time of the observatory, the detection rate is suppressed by the FOV. This suppression is captured by a $\Delta t$-dependent detection efficiency, $ \epsilon(\Delta t)$, obtained by computing the probability that two images fall within the observation time window (see SUPP for the detailed computation). Another key parameter is the system-equivalent flux density (SEFD), which sets the minimum detectable flux $S_{\text{min}} = \text{SEFD}/\sqrt{\Delta T \Delta f }$, where $\Delta T$ is the typical time duration of an FRB burst and $\Delta f$ is the bandwidth of the burst.

Coherent all-sky monitors~\cite{Connor:2022bwl}, such as Bustling Universe Radio Survey Telescope in Taiwan (BURSTT) \cite{Lin:2022wgp} and the Square Kilometer Array Phase 2 (SKA2)~\cite{braun2019}, are well suited for detecting lensed FRBs with long time delays. We also consider the Canadian Hydrogen Intensity Mapping Experiment
 (CHIME)~\cite{Amiri_2018}, projecting its sensitivity for a 20-year observation period. A similar analysis can also be applied to other all-sky monitors, such as future extensions of the FAST array and the Cosmic Antennae~\cite{Normile2025BURSTT,2024Jiang}.  
 
 The parameters for each observatory are summarized in~\tabref{observatories}.  CHIME has an SEFD of 80--90 Jy ($S_{\min} \approx 0.1\,\text{Jy}$) and a FOV of about $140\,\text{deg}^2$ ($2^\circ$ East-West span)~\cite{Amiri_2018}. About $10^5$ bursts are expected to be observed by CHIME in 20 years. BURSTT has a FOV of 5000 $\text{deg}^2$, spanning roughly $70^\circ$ in the East-West direction~\cite{Lin:2022wgp}, yielding $\epsilon(\Delta t) \approx 0.015$ for long time delays. Its SEFD of 600 Jy corresponds to $S_{\text{min}} \approx 0.94$ Jy over a bandwidth of about 400 MHz~\cite{Lin:2022wgp}, and approximately $5 \times 10^5$ bursts are expected within a 10-year period. SKA2 achieves a significantly better SEFD, leading to a much larger number of expected events, but with a smaller FOV of about $100$--$150\,\text{deg}^2$ ($\sim 10^\circ$ East-West span)~\cite{braun2019}. SKA2 operates in two frequency bands: SKA2-Low (50--350 MHz) and SKA2-Mid (0.35--15.4 GHz), which we assume to have approximately the same FOVs. SKA2-Low has an SEFD of 0.2--0.3 Jy ($S_{\min} \approx 4 \times 10^{-4}\,\text{Jy}$), while SKA2-Mid has an SEFD of 0.1--0.2 Jy ($S_{\min} \approx 10^{-4}\,\text{Jy}$)~\cite{braun2019}. Approximately $10^6-10^7$ bursts are expected to be observed by SKA2 within ten years.

\textit{(3) Lensing event identification}. Strong lensing of FRBs has not yet been identified, although several searches have targeted time-delay signatures from gravitational lensing by MACHOs~\cite{Munoz:2016tmg}. MACHOs are modeled as point-like lenses uniformly distributed throughout the universe, which are different from halos.

A potential source of false positives in FRB lensing searches is repetition from intrinsically active FRB sources, since repeated bursts from the same source may in some cases may appear similar. Because the physical origin of FRB emission is still not well understood, the corresponding false-positive rate is presently difficult to quantify. In this work, we assume that such contamination is negligible. Any nonzero false-positive contribution would only weaken the resulting constraints on SIDM models, so our assumption is conservative. To distinguish genuine lensing events from intrinsic repetition, one can rely on waveform fitting \cite{CHIMEFRB:2022xzl,Kader:2024uqm,Xiong:2025gtw}, exploiting the fact that multiple lensed images arise from the same burst and should therefore share the same underlying signal, up to an overall magnification factor and a fixed time delay. In practice, this similarity can be tested through consistency in the pulse morphology, dynamic spectrum, and sub-burst structure. Under this assumption, we take the identification efficiency for true lensed events to be close to 
100\% and leave the detailed study to the future. In addition, actively repeating FRB sources may themselves enhance lensing identification, since a lensed repeater would produce duplicated burst sequences with a common relative time delay and identical waveform between image pairs, yielding a significantly cleaner and higher signal-to-noise signature than in the non-repeating case.

\begin{table}[t]
    \centering
        \captionsetup{position=top}
        \caption{Relevant performance parameters of the observatories considered in this work. The fourth and fifth columns list the expected number of detected FRB events over the next 10 years (20 years for CHIME) and the corresponding detection efficiency, respectively.}
    \begin{tabular}{l c c c c}
       \hline
         Obs. & $S_{\text{min}} [\text{Jy}]$ & FOV [deg${}^2$]& $N_{\text{exp}}$ & $\epsilon(\Delta t)$ \\ \hline
                CHIME  & 0.1 & 140 &  $ \sim 10^5$  & $\sim  10^{-5}$\\
       BURSTT  & 0.94 & 5000 &  $\sim 5\cdot10^5$ & $\sim 0.015$ \\ 
       SKA2-Low  & $4\cdot 10^{-4}$&  $\sim 100$ &  $ \sim 4\cdot10^6$  & $\gtrsim 6 \cdot 10^{-5}$\\   
       SKA2-Mid  & $10^{-4}$ & $\sim 100$ &  $ \gtrsim8 \cdot 10^6$ & $\sim6 \cdot 10^{-5}$\\
    \hline 
    \end{tabular}
    \label{tab:observatories}
\end{table}

\begin{figure*}
    \centering        \includegraphics[width=0.96\linewidth]{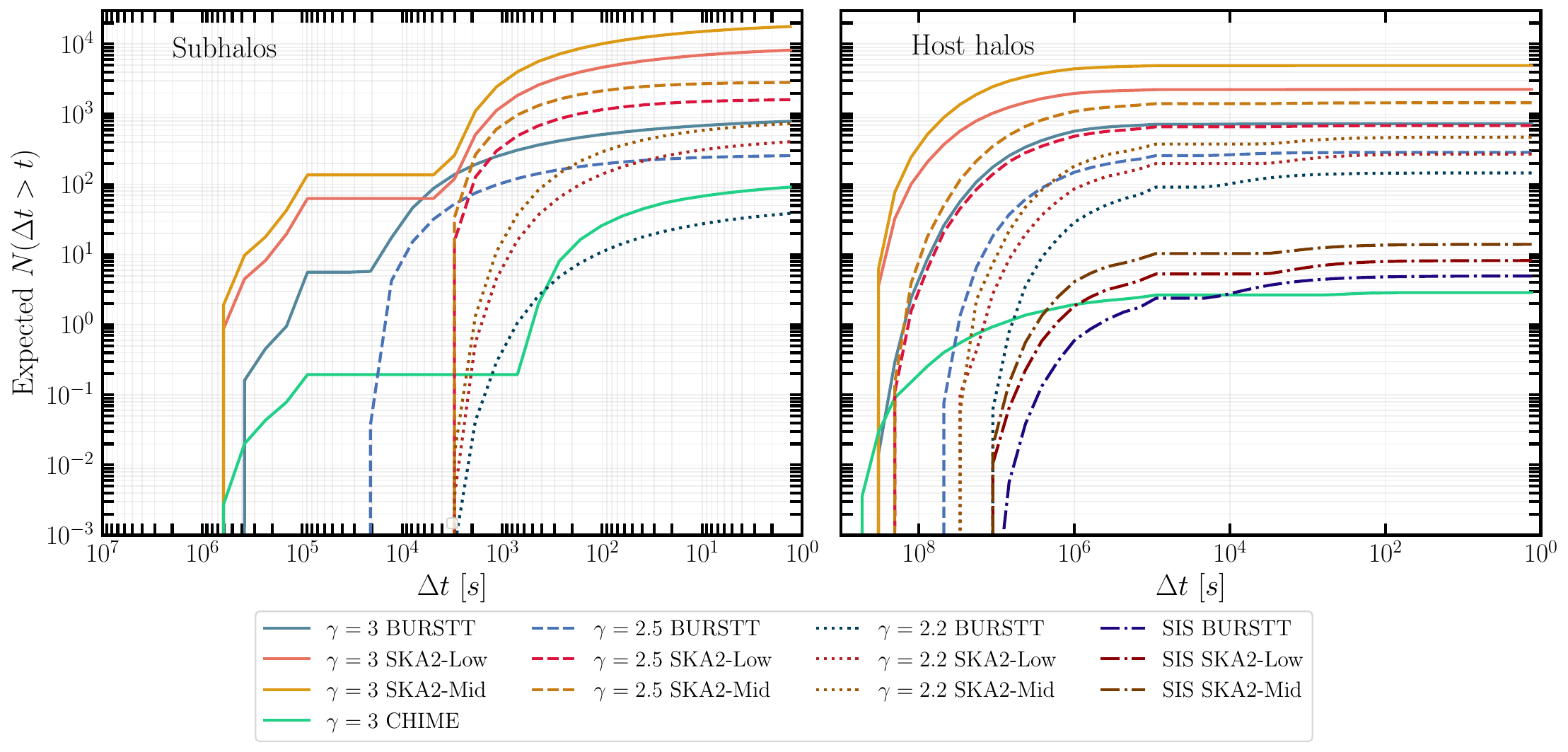}
    \caption{(\emph{Left}) The expected cumulative distribution of time delays for FRB images lensed by SIDM subhalos for various subhalo profiles and observatories, assuming all of the subhalos underwent core-collapse. (\emph{Right}) The expected cumulative distribution of time delays for FRB images lensed by SIDM host halos that underwent core-collapse.}
    \label{fig:delay}
\end{figure*}

\noindent \textbf{\textit{Results.}} We compute the total lensing rate of FRBs by core-collapsed (sub)halos by integrating the optical depth over the source distribution,
\begin{equation}
    N_{\text{tot}} = \int dz_s \int dL_s \frac{d^2n_s}{dz_s dL_s}  (1-e^{-\tau(z_s,L_s)}),
    \label{eq:tot}
\end{equation}
where the FRB source distribution ${d^2n_s}/{dz_s dL_s}$ is obtained from the cumulative redshift-luminosity distribution discussed earlier. We work in the single-lensing regime, in which each FRB is lensed at most once along its line of sight, and expand the exponential in~\eqref{tot} to first order. The optical depth for an FRB at $z_s$ is then
\begin{align}
    \tau(z_s, L_s) =&{} \int 2\pi\beta d \beta \int  dM_l \int  dz_l   \frac{d^2n_l}{dz_l dM_l} \epsilon(\Delta t)  \nonumber\\
    & \Theta \left[ \min_i( \mu_i L_s /(4\pi D_s^2 (1+z_s)^4))  -S_{\text{min}} \right],
    \label{eq:opt}
\end{align}
 where ${d^2n_l}/{dz_l dM_l}$ is the mass and redshift distribution of the lensing halos, $\mu_i$ is the amplification factor of the $i$-th image,  
  and $D_s$ denotes the angular diameter distance to the source, determined by $z_s$. Detailed expressions and integration limits are provided in the SUPP. We first evaluate~\eqref{opt} and substitute the result into~\eqref{tot} to obtain the total number of lensing events. We then generate
a set of mock lensing events by sampling the parameter set $\{M_l,z_s,z_l,L_s, \beta \}$ from the integrands of~(\ref{eq:tot}) and~(\ref{eq:opt})  and computing the time delay event by event. 

 \begin{figure}
    \centering    \includegraphics[width=0.99\linewidth]{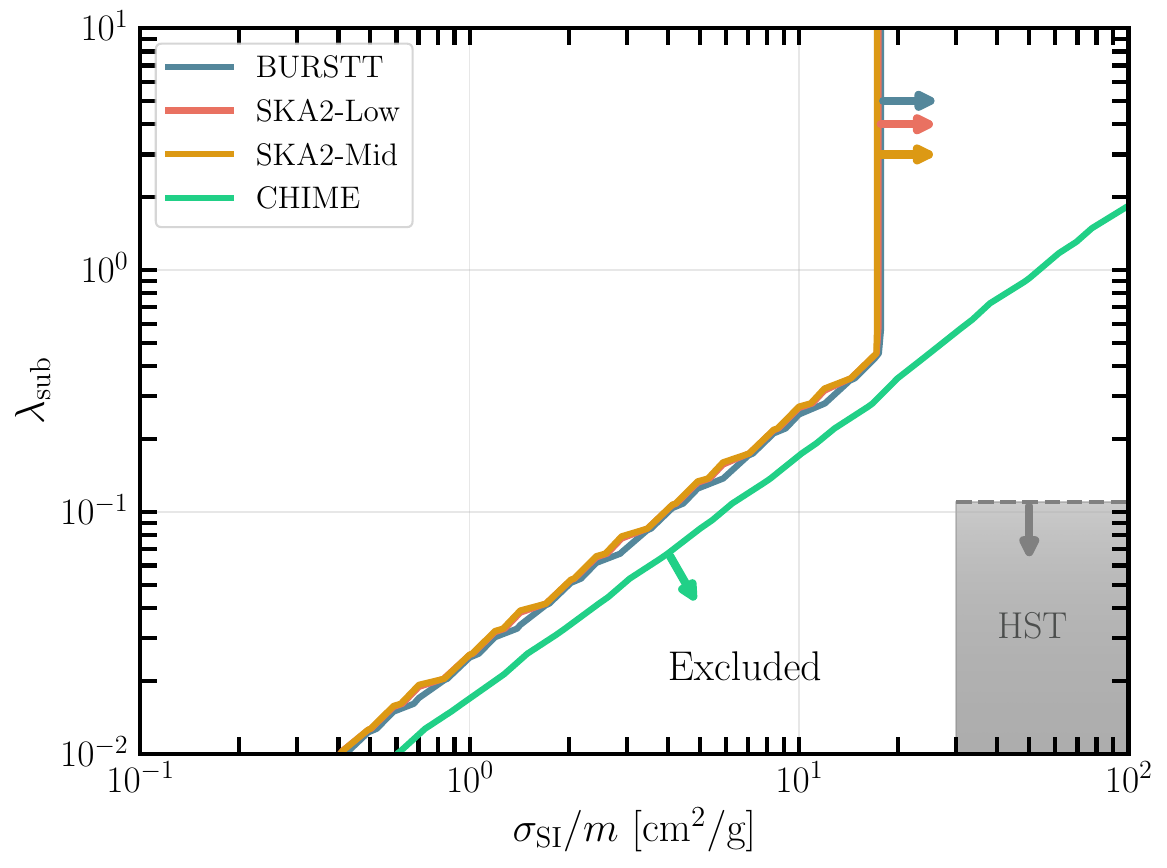}
    \caption{Projected 95\% CL exclusion regions in the $\lambda_{\text{sub}}$--$\sigma_{\text{SI}}/m$ parameter space, derived under the assumption that no excess beyond singular isothermal spheres lens model is observed. Different colors correspond to constraints from different observatories. The gray shaded region is excluded by an analysis of 11 quadruply-imaged quasars observed with the Hubble Space Telescope (HST)~\cite{Gilman:2022ida}.  }
    \label{fig:proj}
\end{figure}

We consider two lensing scenarios: (A) an FRB is lensed by a core-collapsed subhalo, and (B) an FRB is lensed by a core-collapsed host halo. The left and right panels of~\figref{delay} show the resulting $\Delta t$ distributions for scenarios (A) and (B), respectively, where the lenses follow either the profile in~(\ref{eq:prof}) or an SIS profile. The blue, red, orange, and green lines indicate the expected number of events for BURSTT, SKA2-Low, SKA2-Mid, and CHIME, respectively.

As the inner slope $\gamma$ increases, the lensing event rate grows and the maximum time delay shifts to larger values accordingly. The cutoffs at time delays, for those shorter than the observational duration (20 yrs for CHIME and 10 yrs for others), arise primarily from the lower end of the lensed halo mass range and the density profiles considered. For the host halo with profile~(\ref{eq:prof}) and $\gamma = 3$, the time delays are instead cut off at the observational duration. The differences in time-delay cutoffs among observatories reflect their different detection efficiencies and flux thresholds. The differences between scenarios (A) and (B) arise primarily from the differences in lensed halo mass ranges and distributions between the subhalo and host halo populations.

We now project the sensitivity of future FRB lensing observations to SIDM parameters. The core collapse of an isolated halo with elastic self-interactions occurs on a timescale $t_c \sim \mathcal{O}(100)\, t_0$, where $t_0$ is the fiducial timescale~\cite{Essig:2018pzq, Gilman:2022ida},
\begin{align}
\nonumber
    t_0 = \left(\frac{1\,\text{cm}^2\text{g}^{-1}}{\sigma_\text{SI}/m}\right) \left(\frac{100\, \text{km s}^{-1}}{v_{\text{max}}}\right) \left( \frac{10^7 M_\odot \text{kpc}^{-3}}{\rho_s}\right) \text{Gyr},
\end{align}
where $v_{\text{max}} = 1.65\sqrt{G \rho_s r_s^2}$ is the maximum circular velocity of the (sub)halo, $G$ is the gravitational constant, and $\sigma_\text{SI}/m$ is the elastic self-interaction cross section strength. We assume a velocity-independent cross section throughout; the extension to velocity-dependent SIDM is straightforward via effective cross sections~\cite{Yang:2022zkd, Yang:2022hkm, Outmezguine:2022bhq}. We define the core-collapse timescales for host halos and subhalos as
\begin{equation}
    t_{c, \text{host}} \equiv \lambda_{\text{field}} t_0, \qquad
    t_{c, \text{sub}}  \equiv \lambda_{\text{sub}}\lambda_{\text{field}} t_0,
\end{equation}
where $\lambda_{\text{field}}$ and $\lambda_{\text{sub}}$ encode the properties of self-interactions and environments for the host and subhalos, respectively. For an isolated SIDM halo with an NFW initial profile, N-body simulations give a collapse time of $t_c \approx 264\, t_0$~\cite{Yang:2023jwn}, so we fix $\lambda_\text{field} = 264$. We note that a compact central baryon distribution can shorten the collapse time below $264\, t_0$~\cite{Zhong:2023yzk}, while repeated baryonic feedback can smooth the central potential and suppress the accelerated collapse. For subhalos, tidal stripping can further accelerate core collapse relative to the isolated case ($\lambda_\text{sub} < 1$), whereas tidal heating and evaporation can delay it ($\lambda_\text{sub} > 1$) (see, e.g.,~\cite{Zeng:2021ldo}). We defer a detailed modeling of $\lambda_\text{sub}$ to future work and treat it as a free parameter with a flat prior over the range $[10^{-2},\, 10]$. Following~\cite{Gilman:2022ida}, we assign the collapse probability for a population of subhalos as
\begin{equation}
    \mathcal{P}_c = \frac{1}{2}\left[1+\tanh\left(\frac{t_{\text{evo}}(z)-t_{c, \text{sub}}(z,\sigma_{\text{SI}}/m, \lambda_\text{sub})}{2\, s_{\text{sub}}}\right)\right],
    \label{eq:prob}
\end{equation}
where $t_{\text{evo}}$ is the evolution time of the subhalo since its formation, which we fix at $z=2$, and $s_{\text{sub}}$ characterizes the scatter in collapse timescales across the subhalo population. Following~\cite{Gilman:2022ida}, we set $s_\text{sub} = 1\,\text{Gyr}$. For the host halo, we apply~\eqref{prob} by replacing $t_{c,\text{sub}}$ with $t_{c, \text{host}}(z, \sigma_\text{SI}/m)$ while keeping the same scatter.

We perform parameter inference on $\sigma_{\text{SI}}/m$ and $\lambda_{\text{sub}}$, assuming core-collapsed halos follow profile~(\ref{eq:prof}) with $\gamma = 3$. For subhalos, we neglect contributions from non-collapsed subhalos, since the lensing probability of NFW or cored halos is negligibly small. For host halos, we assume non-collapsed halos follow SIS profiles. Lensed events with time delays in the range $[1,\, 10^9]$\,s are binned into 50 logarithmically spaced bins. The likelihood is constructed as a product of Poisson distributions over bins, $\mathcal{L} = \prod_i n_i(\lambda_{\text{sub}}, \sigma_{\text{SI}}/m)^{N_i}\, e^{-n_i(\lambda_{\text{sub}}, \sigma_{\text{SI}}/m)}/\left({N_i!}\right)$, where $N_i$ is the observed number of lensed events in the $i$-th bin and $n_i$ is the expected number for the given parameters. The 95\% CL exclusion limit is derived from the test statistic $t_{\text{exc}} \equiv -2\ln(\mathcal{L}/\mathcal{L}_{\text{max}})$, where $\mathcal{L}_{\text{max}}$ is the likelihood maximized over the $\{\lambda_{\text{sub}}, \sigma_{\text{SI}}/m\}$ plane, and the exclusion region corresponds to $t_{\text{exc}} \geq 5.99$.

To estimate the time-delay distribution of lensed events, we first compute the expected number counts for subhalo and host halo lensing using~\eqref{tot}, then sample from the integrand using \texttt{emcee}~\cite{Foreman-Mackey:2012any} (see SUPP for details). The samples are subsequently reweighted by~\eqref{prob} to account for the core-collapse condition, which is controlled by $\lambda_{\text{sub}}$ and $\sigma_{\text{SI}}/m$.

 \figref{proj} shows the projected sensitivity of BURSTT, SKA2-Low, SKA2-Mid, and CHIME to the parameters $\{\lambda_{\text{sub}}, \ \sigma_{\text{SI}}/m\}$, where the exclusion region is derived from a mock time-delay dataset generated under the assumption that
no halo has experienced core collapse. We find that BURSTT, SKA2-Low, and SKA2-Mid achieve similar sensitivities when $\lambda_{\text{sub}} \lesssim 0.4$, where subhalo lensing dominates and constraints on $\lambda_{\text{sub}}$ and $\sigma_{\text{SI}}/m$ exclude $(\sigma_{\text{SI}}/m)/\lambda_{\text{sub}} \gtrsim 40\,\text{cm}^2/\text{g}$. When $\lambda_{\text{sub}} \gtrsim 0.4$, host halo lensing dominates and the constraints depend on $\sigma_{\text{SI}}/m$ alone, becoming insensitive to $\lambda_{\text{sub}}$. The overall excluded regions can be summarized as $\sigma_{\text{SI}}/m \gtrsim \min\{18,\, 40\lambda_{\text{sub}}\}\,\text{cm}^2/\text{g}$. CHIME, with 20 years of observation, achieves comparable sensitivity to subhalo lensing, though slightly weaker, but remains insensitive to core collapse of host halos. The degeneracy features between BURSTT, SKA2-Low and SKA2-Mid along the exclusion boundary arise from the threshold-like dependence of~\eqref{prob} on $\{\lambda_{\text{sub}}, \sigma_{\text{SI}}/m\}$, where the number of lensing events rises sharply once the parameters cross the threshold for core collapse. In~\figref{proj}, we also show the sensitivity of the Hubble Space Telescope, derived from 11 quadruply-imaged quasars~\cite{Gilman:2022ida}, translated to our parameter space by replacing the velocity-dependent cross section with the effective cross section and interpolating their constraints from benchmark models 1, 2, and 5.

\noindent \textbf{\textit{Conclusions.}} We have demonstrated that strong lensing of FRBs provides a powerful probe of SIDM via time delays. Modeling core-collapsed halos using a cored power-law density profile with $2 \leq \gamma \leq 3$, we showed that these ultra-dense structures dramatically enhance the lensing cross section and produce significantly longer time delays compared to NFW or SIS profiles typical of CDM halos. Current and future all-sky monitors including CHIME, BURSTT, SKA2-Low, and SKA2-Mid are expected to detect $10^5$ to $10^7$ FRBs over the next decade, providing substantial constraining power over SIDM parameters $\{\lambda_{\text{sub}}, \ \sigma_{\text{SI}}/m\}$.  

The analysis framework developed here can be readily extended to other scenarios producing ultra-dense small-scale structures, such as core-collapsed globular clusters~\cite{spitzer2014dynamical} or primordial black hole clusters. It will also be interesting in future work to extend this framework to strongly lensed gamma-ray bursts and supernovae, whose transient light curves provide complementary short and long timescale time-delay observables and have already been recognized as promising lensing probes~\cite{Grossman:1994je,Ji:2018rvg,Goobar:2024dzh,Levan:2025ool,Suyu:2023jue}. Looking ahead, multi-wavelength follow-up observations that identify FRB host galaxies and resolve lensing systems in the optical or infrared will enable direct confirmation of lensing events, improved characterization of lens properties, and, in particular, better discrimination between intrinsic repetitions and genuinely lensed images.

\section*{Acknowledgment}
We thank T.K. Chan, Daniel Gilman, Jeremy Lim, Huangyu Xiao, and Bing Zhang for useful discussions. Y.H. and Y.Z. are supported by the GRF Grants No. 11302824 and No. 11310925 from the Research Grants Council, University Grants Committee, and the Grants No. 9610645 and No. 7020130 from the City University of Hong Kong. C.Z. is supported by the Fundamental Research Funds for the Central Universities and the Jockey Club Institute for
Advanced Study at the Hong Kong University of Science
and Technology.
W.-Y. W. acknowledges support from the NSFC (No.12261141690 and No.12403058), and the Strategic Priority Research Program of the CAS (No. XDB0550300). Y.H. also thanks the Mainz Institute for Theoretical Physics (MITP) of the PRISMA+ Cluster of Excellence
(Project ID 390831469) for its hospitality and partial support during the completion of this work. Y.Z. acknowledges the Aspen Center for Physics, which is supported by NSF Grant No. PHY-2210452, for their hospitality during the completion of this study.

\bibliography{bib.bib}

%merlin.mbs apsrev4-1.bst 2010-07-25 4.21a (PWD, AO, DPC) hacked
%Control: key (0)
%Control: author (72) initials jnrlst
%Control: editor formatted (1) identically to author
%Control: production of article title (-1) disabled
%Control: page (0) single
%Control: year (1) truncated
%Control: production of eprint (0) enabled
\begin{thebibliography}{84}%
\makeatletter
\providecommand \@ifxundefined [1]{%
 \@ifx{#1\undefined}
}%
\providecommand \@ifnum [1]{%
 \ifnum #1\expandafter \@firstoftwo
 \else \expandafter \@secondoftwo
 \fi
}%
\providecommand \@ifx [1]{%
 \ifx #1\expandafter \@firstoftwo
 \else \expandafter \@secondoftwo
 \fi
}%
\providecommand \natexlab [1]{#1}%
\providecommand \enquote  [1]{``#1''}%
\providecommand \bibnamefont  [1]{#1}%
\providecommand \bibfnamefont [1]{#1}%
\providecommand \citenamefont [1]{#1}%
\providecommand \href@noop [0]{\@secondoftwo}%
\providecommand \href [0]{\begingroup \@sanitize@url \@href}%
\providecommand \@href[1]{\@@startlink{#1}\@@href}%
\providecommand \@@href[1]{\endgroup#1\@@endlink}%
\providecommand \@sanitize@url [0]{\catcode `\\12\catcode `\$12\catcode
  `\&12\catcode `\#12\catcode `\^12\catcode `\_12\catcode `\%12\relax}%
\providecommand \@@startlink[1]{}%
\providecommand \@@endlink[0]{}%
\providecommand \url  [0]{\begingroup\@sanitize@url \@url }%
\providecommand \@url [1]{\endgroup\@href {#1}{\urlprefix }}%
\providecommand \urlprefix  [0]{URL }%
\providecommand \Eprint [0]{\href }%
\providecommand \doibase [0]{http://dx.doi.org/}%
\providecommand \selectlanguage [0]{\@gobble}%
\providecommand \bibinfo  [0]{\@secondoftwo}%
\providecommand \bibfield  [0]{\@secondoftwo}%
\providecommand \translation [1]{[#1]}%
\providecommand \BibitemOpen [0]{}%
\providecommand \bibitemStop [0]{}%
\providecommand \bibitemNoStop [0]{.\EOS\space}%
\providecommand \EOS [0]{\spacefactor3000\relax}%
\providecommand \BibitemShut  [1]{\csname bibitem#1\endcsname}%
\let\auto@bib@innerbib\@empty
%</preamble>
\bibitem [{\citenamefont {Aghanim}\ \emph {et~al.}(2020)\citenamefont {Aghanim}
  \emph {et~al.}}]{Planck:2018vyg}%
  \BibitemOpen
  \bibfield  {author} {\bibinfo {author} {\bibfnamefont {N.}~\bibnamefont
  {Aghanim}} \emph {et~al.} (\bibinfo {collaboration} {Planck}),\ }\href
  {\doibase 10.1051/0004-6361/201833910} {\bibfield  {journal} {\bibinfo
  {journal} {Astron. Astrophys.}\ }\textbf {\bibinfo {volume} {641}},\ \bibinfo
  {pages} {A6} (\bibinfo {year} {2020})},\ \bibinfo {note} {[Erratum:
  Astron.Astrophys. 652, C4 (2021)]},\ \Eprint
  {http://arxiv.org/abs/1807.06209} {arXiv:1807.06209 [astro-ph.CO]}
  \BibitemShut {NoStop}%
\bibitem [{\citenamefont {Bullock}\ and\ \citenamefont
  {Boylan-Kolchin}(2017)}]{Bullock:2017xww}%
  \BibitemOpen
  \bibfield  {author} {\bibinfo {author} {\bibfnamefont {J.~S.}\ \bibnamefont
  {Bullock}}\ and\ \bibinfo {author} {\bibfnamefont {M.}~\bibnamefont
  {Boylan-Kolchin}},\ }\href {\doibase 10.1146/annurev-astro-091916-055313}
  {\bibfield  {journal} {\bibinfo  {journal} {Ann. Rev. Astron. Astrophys.}\
  }\textbf {\bibinfo {volume} {55}},\ \bibinfo {pages} {343} (\bibinfo {year}
  {2017})},\ \Eprint {http://arxiv.org/abs/1707.04256} {arXiv:1707.04256
  [astro-ph.CO]} \BibitemShut {NoStop}%
\bibitem [{\citenamefont {Tulin}\ and\ \citenamefont
  {Yu}(2018)}]{Tulin:2017ara}%
  \BibitemOpen
  \bibfield  {author} {\bibinfo {author} {\bibfnamefont {S.}~\bibnamefont
  {Tulin}}\ and\ \bibinfo {author} {\bibfnamefont {H.-B.}\ \bibnamefont {Yu}},\
  }\href {\doibase 10.1016/j.physrep.2017.11.004} {\bibfield  {journal}
  {\bibinfo  {journal} {Phys. Rept.}\ }\textbf {\bibinfo {volume} {730}},\
  \bibinfo {pages} {1} (\bibinfo {year} {2018})},\ \Eprint
  {http://arxiv.org/abs/1705.02358} {arXiv:1705.02358 [hep-ph]} \BibitemShut
  {NoStop}%
\bibitem [{\citenamefont {Dutra}\ \emph {et~al.}(2025)\citenamefont {Dutra},
  \citenamefont {Natarajan},\ and\ \citenamefont {Gilman}}]{Dutra:2024qac}%
  \BibitemOpen
  \bibfield  {author} {\bibinfo {author} {\bibfnamefont {I.}~\bibnamefont
  {Dutra}}, \bibinfo {author} {\bibfnamefont {P.}~\bibnamefont {Natarajan}}, \
  and\ \bibinfo {author} {\bibfnamefont {D.}~\bibnamefont {Gilman}},\ }\href
  {\doibase 10.3847/1538-4357/ad9b09} {\bibfield  {journal} {\bibinfo
  {journal} {Astrophys. J.}\ }\textbf {\bibinfo {volume} {978}},\ \bibinfo
  {pages} {38} (\bibinfo {year} {2025})},\ \Eprint
  {http://arxiv.org/abs/2406.17024} {arXiv:2406.17024 [astro-ph.CO]}
  \BibitemShut {NoStop}%
\bibitem [{\citenamefont {Powell}\ \emph {et~al.}(2025)\citenamefont {Powell},
  \citenamefont {McKean}, \citenamefont {Vegetti}, \citenamefont {Spingola},
  \citenamefont {White},\ and\ \citenamefont {Fassnacht}}]{Powell:2025rmj}%
  \BibitemOpen
  \bibfield  {author} {\bibinfo {author} {\bibfnamefont {D.~M.}\ \bibnamefont
  {Powell}}, \bibinfo {author} {\bibfnamefont {J.~P.}\ \bibnamefont {McKean}},
  \bibinfo {author} {\bibfnamefont {S.}~\bibnamefont {Vegetti}}, \bibinfo
  {author} {\bibfnamefont {C.}~\bibnamefont {Spingola}}, \bibinfo {author}
  {\bibfnamefont {S.~D.~M.}\ \bibnamefont {White}}, \ and\ \bibinfo {author}
  {\bibfnamefont {C.~D.}\ \bibnamefont {Fassnacht}},\ }\href {\doibase
  10.1038/s41550-025-02651-2} {\bibfield  {journal} {\bibinfo  {journal}
  {Nature Astron.}\ }\textbf {\bibinfo {volume} {9}},\ \bibinfo {pages} {1714}
  (\bibinfo {year} {2025})},\ \Eprint {http://arxiv.org/abs/2510.07382}
  {arXiv:2510.07382 [astro-ph.CO]} \BibitemShut {NoStop}%
\bibitem [{\citenamefont {Tajalli}\ \emph {et~al.}(2025)\citenamefont
  {Tajalli}, \citenamefont {Vegetti}, \citenamefont {O'Riordan}, \citenamefont
  {White}, \citenamefont {Fassnacht}, \citenamefont {Powell}, \citenamefont
  {McKean},\ and\ \citenamefont {Despali}}]{Tajalli:2025qjx}%
  \BibitemOpen
  \bibfield  {author} {\bibinfo {author} {\bibfnamefont {M.}~\bibnamefont
  {Tajalli}}, \bibinfo {author} {\bibfnamefont {S.}~\bibnamefont {Vegetti}},
  \bibinfo {author} {\bibfnamefont {C.~M.}\ \bibnamefont {O'Riordan}}, \bibinfo
  {author} {\bibfnamefont {S.~D.~M.}\ \bibnamefont {White}}, \bibinfo {author}
  {\bibfnamefont {C.~D.}\ \bibnamefont {Fassnacht}}, \bibinfo {author}
  {\bibfnamefont {D.~M.}\ \bibnamefont {Powell}}, \bibinfo {author}
  {\bibfnamefont {J.~P.}\ \bibnamefont {McKean}}, \ and\ \bibinfo {author}
  {\bibfnamefont {G.}~\bibnamefont {Despali}},\ }\href {\doibase
  10.1093/mnras/staf1357} {\bibfield  {journal} {\bibinfo  {journal} {Mon. Not.
  Roy. Astron. Soc.}\ }\textbf {\bibinfo {volume} {543}},\ \bibinfo {pages}
  {540} (\bibinfo {year} {2025})},\ \Eprint {http://arxiv.org/abs/2505.07944}
  {arXiv:2505.07944 [astro-ph.CO]} \BibitemShut {NoStop}%
\bibitem [{\citenamefont {Enzi}\ \emph {et~al.}(2025)\citenamefont {Enzi},
  \citenamefont {Krawczyk}, \citenamefont {Ballard},\ and\ \citenamefont
  {Collett}}]{Enzi2025}%
  \BibitemOpen
  \bibfield  {author} {\bibinfo {author} {\bibfnamefont {W.~J.~R.}\
  \bibnamefont {Enzi}}, \bibinfo {author} {\bibfnamefont {C.~M.}\ \bibnamefont
  {Krawczyk}}, \bibinfo {author} {\bibfnamefont {D.~J.}\ \bibnamefont
  {Ballard}}, \ and\ \bibinfo {author} {\bibfnamefont {T.~E.}\ \bibnamefont
  {Collett}},\ }\href {\doibase 10.1093/mnras/staf697} {\bibfield  {journal}
  {\bibinfo  {journal} {Monthly Notices of the Royal Astronomical Society}\
  }\textbf {\bibinfo {volume} {540}},\ \bibinfo {pages} {247} (\bibinfo {year}
  {2025})},\ \Eprint
  {http://arxiv.org/abs/https://academic.oup.com/mnras/article-pdf/540/1/247/63037501/staf697.pdf}
  {https://academic.oup.com/mnras/article-pdf/540/1/247/63037501/staf697.pdf}
  \BibitemShut {NoStop}%
\bibitem [{\citenamefont {He}\ \emph {et~al.}(2025)\citenamefont {He} \emph
  {et~al.}}]{He:2025wco}%
  \BibitemOpen
  \bibfield  {author} {\bibinfo {author} {\bibfnamefont {Q.}~\bibnamefont {He}}
  \emph {et~al.},\ }\href {\doibase 10.3847/2041-8213/ae072d} {\bibfield
  {journal} {\bibinfo  {journal} {Astrophys. J. Lett.}\ }\textbf {\bibinfo
  {volume} {991}},\ \bibinfo {pages} {L53} (\bibinfo {year} {2025})},\ \Eprint
  {http://arxiv.org/abs/2506.07978} {arXiv:2506.07978 [astro-ph.CO]}
  \BibitemShut {NoStop}%
\bibitem [{\citenamefont {Minor}\ \emph {et~al.}(2021)\citenamefont {Minor},
  \citenamefont {Gad-Nasr}, \citenamefont {Kaplinghat},\ and\ \citenamefont
  {Vegetti}}]{Minor:2020hic}%
  \BibitemOpen
  \bibfield  {author} {\bibinfo {author} {\bibfnamefont {Q.~E.}\ \bibnamefont
  {Minor}}, \bibinfo {author} {\bibfnamefont {S.}~\bibnamefont {Gad-Nasr}},
  \bibinfo {author} {\bibfnamefont {M.}~\bibnamefont {Kaplinghat}}, \ and\
  \bibinfo {author} {\bibfnamefont {S.}~\bibnamefont {Vegetti}},\ }\href
  {\doibase 10.1093/mnras/stab2247} {\bibfield  {journal} {\bibinfo  {journal}
  {Mon. Not. Roy. Astron. Soc.}\ }\textbf {\bibinfo {volume} {507}},\ \bibinfo
  {pages} {1662} (\bibinfo {year} {2021})},\ \Eprint
  {http://arxiv.org/abs/2011.10627} {arXiv:2011.10627 [astro-ph.GA]}
  \BibitemShut {NoStop}%
\bibitem [{\citenamefont {Vegetti}\ \emph {et~al.}(2010)\citenamefont
  {Vegetti}, \citenamefont {Koopmans}, \citenamefont {Bolton}, \citenamefont
  {Treu},\ and\ \citenamefont {Gavazzi}}]{Vegetti2010}%
  \BibitemOpen
  \bibfield  {author} {\bibinfo {author} {\bibfnamefont {S.}~\bibnamefont
  {Vegetti}}, \bibinfo {author} {\bibfnamefont {L.~V.~E.}\ \bibnamefont
  {Koopmans}}, \bibinfo {author} {\bibfnamefont {A.}~\bibnamefont {Bolton}},
  \bibinfo {author} {\bibfnamefont {T.}~\bibnamefont {Treu}}, \ and\ \bibinfo
  {author} {\bibfnamefont {R.}~\bibnamefont {Gavazzi}},\ }\href {\doibase
  10.1111/j.1365-2966.2010.16865.x} {\bibfield  {journal} {\bibinfo  {journal}
  {Monthly Notices of the Royal Astronomical Society}\ }\textbf {\bibinfo
  {volume} {408}},\ \bibinfo {pages} {1969} (\bibinfo {year} {2010})},\ \Eprint
  {http://arxiv.org/abs/https://academic.oup.com/mnras/article-pdf/408/4/1969/4219080/mnras0408-1969.pdf}
  {https://academic.oup.com/mnras/article-pdf/408/4/1969/4219080/mnras0408-1969.pdf}
  \BibitemShut {NoStop}%
\bibitem [{\citenamefont {{Vegetti}}\ \emph {et~al.}(2026)\citenamefont
  {{Vegetti}}, \citenamefont {{White}}, \citenamefont {{McKean}}, \citenamefont
  {{Powell}}, \citenamefont {{Spingola}}, \citenamefont {{Massari}},
  \citenamefont {{Despali}},\ and\ \citenamefont
  {{Fassnacht}}}]{Vegetti:2026mmx}%
  \BibitemOpen
  \bibfield  {author} {\bibinfo {author} {\bibfnamefont {S.}~\bibnamefont
  {{Vegetti}}}, \bibinfo {author} {\bibfnamefont {S.~D.~M.}\ \bibnamefont
  {{White}}}, \bibinfo {author} {\bibfnamefont {J.~P.}\ \bibnamefont
  {{McKean}}}, \bibinfo {author} {\bibfnamefont {D.~M.}\ \bibnamefont
  {{Powell}}}, \bibinfo {author} {\bibfnamefont {C.}~\bibnamefont
  {{Spingola}}}, \bibinfo {author} {\bibfnamefont {D.}~\bibnamefont
  {{Massari}}}, \bibinfo {author} {\bibfnamefont {G.}~\bibnamefont
  {{Despali}}}, \ and\ \bibinfo {author} {\bibfnamefont {C.~D.}\ \bibnamefont
  {{Fassnacht}}},\ }\href {\doibase 10.1038/s41550-025-02746-w} {\bibfield
  {journal} {\bibinfo  {journal} {Nature Astronomy}\ } (\bibinfo {year}
  {2026}),\ 10.1038/s41550-025-02746-w},\ \Eprint
  {http://arxiv.org/abs/2601.02466} {arXiv:2601.02466 [astro-ph.CO]}
  \BibitemShut {NoStop}%
\bibitem [{\citenamefont {Navarro}\ \emph {et~al.}(1996)\citenamefont
  {Navarro}, \citenamefont {Frenk},\ and\ \citenamefont
  {White}}]{Navarro:1995iw}%
  \BibitemOpen
  \bibfield  {author} {\bibinfo {author} {\bibfnamefont {J.~F.}\ \bibnamefont
  {Navarro}}, \bibinfo {author} {\bibfnamefont {C.~S.}\ \bibnamefont {Frenk}},
  \ and\ \bibinfo {author} {\bibfnamefont {S.~D.~M.}\ \bibnamefont {White}},\
  }\href {\doibase 10.1086/177173} {\bibfield  {journal} {\bibinfo  {journal}
  {Astrophys. J.}\ }\textbf {\bibinfo {volume} {462}},\ \bibinfo {pages} {563}
  (\bibinfo {year} {1996})},\ \Eprint {http://arxiv.org/abs/astro-ph/9508025}
  {arXiv:astro-ph/9508025} \BibitemShut {NoStop}%
\bibitem [{\citenamefont {Navarro}\ \emph {et~al.}(1997)\citenamefont
  {Navarro}, \citenamefont {Frenk},\ and\ \citenamefont
  {White}}]{Navarro:1996gj}%
  \BibitemOpen
  \bibfield  {author} {\bibinfo {author} {\bibfnamefont {J.~F.}\ \bibnamefont
  {Navarro}}, \bibinfo {author} {\bibfnamefont {C.~S.}\ \bibnamefont {Frenk}},
  \ and\ \bibinfo {author} {\bibfnamefont {S.~D.~M.}\ \bibnamefont {White}},\
  }\href {\doibase 10.1086/304888} {\bibfield  {journal} {\bibinfo  {journal}
  {Astrophys. J.}\ }\textbf {\bibinfo {volume} {490}},\ \bibinfo {pages} {493}
  (\bibinfo {year} {1997})},\ \Eprint {http://arxiv.org/abs/astro-ph/9611107}
  {arXiv:astro-ph/9611107} \BibitemShut {NoStop}%
\bibitem [{\citenamefont {Essig}\ \emph {et~al.}(2019)\citenamefont {Essig},
  \citenamefont {Mcdermott}, \citenamefont {Yu},\ and\ \citenamefont
  {Zhong}}]{Essig:2018pzq}%
  \BibitemOpen
  \bibfield  {author} {\bibinfo {author} {\bibfnamefont {R.}~\bibnamefont
  {Essig}}, \bibinfo {author} {\bibfnamefont {S.~D.}\ \bibnamefont
  {Mcdermott}}, \bibinfo {author} {\bibfnamefont {H.-B.}\ \bibnamefont {Yu}}, \
  and\ \bibinfo {author} {\bibfnamefont {Y.-M.}\ \bibnamefont {Zhong}},\ }\href
  {\doibase 10.1103/PhysRevLett.123.121102} {\bibfield  {journal} {\bibinfo
  {journal} {Phys. Rev. Lett.}\ }\textbf {\bibinfo {volume} {123}},\ \bibinfo
  {pages} {121102} (\bibinfo {year} {2019})},\ \Eprint
  {http://arxiv.org/abs/1809.01144} {arXiv:1809.01144 [hep-ph]} \BibitemShut
  {NoStop}%
\bibitem [{\citenamefont {Gurian}\ and\ \citenamefont
  {May}(2025)}]{Gurian:2025zpc}%
  \BibitemOpen
  \bibfield  {author} {\bibinfo {author} {\bibfnamefont {J.}~\bibnamefont
  {Gurian}}\ and\ \bibinfo {author} {\bibfnamefont {S.}~\bibnamefont {May}},\
  }\href {\doibase 10.1103/2ycz-3fvv} {\bibfield  {journal} {\bibinfo
  {journal} {Phys. Rev. Lett.}\ }\textbf {\bibinfo {volume} {135}},\ \bibinfo
  {pages} {221001} (\bibinfo {year} {2025})},\ \Eprint
  {http://arxiv.org/abs/2505.15903} {arXiv:2505.15903 [astro-ph.CO]}
  \BibitemShut {NoStop}%
\bibitem [{\citenamefont {Fischer}\ \emph {et~al.}(2025)\citenamefont
  {Fischer}, \citenamefont {Yu},\ and\ \citenamefont
  {Dolag}}]{Fischer:2025rky}%
  \BibitemOpen
  \bibfield  {author} {\bibinfo {author} {\bibfnamefont {M.~S.}\ \bibnamefont
  {Fischer}}, \bibinfo {author} {\bibfnamefont {H.-B.}\ \bibnamefont {Yu}}, \
  and\ \bibinfo {author} {\bibfnamefont {K.}~\bibnamefont {Dolag}},\ }\href
  {\doibase 10.1051/0004-6361/202556189} {\bibfield  {journal} {\bibinfo
  {journal} {Astron. Astrophys.}\ }\textbf {\bibinfo {volume} {703}},\ \bibinfo
  {pages} {A234} (\bibinfo {year} {2025})},\ \Eprint
  {http://arxiv.org/abs/2506.06269} {arXiv:2506.06269 [astro-ph.CO]}
  \BibitemShut {NoStop}%
\bibitem [{\citenamefont {Zeng}\ \emph {et~al.}(2025)\citenamefont {Zeng},
  \citenamefont {Peter}, \citenamefont {Du}, \citenamefont {Yang},
  \citenamefont {Benson}, \citenamefont {Cyr-Racine}, \citenamefont {Jiang},
  \citenamefont {Mace},\ and\ \citenamefont {Metcalf}}]{Zeng:2023fnj}%
  \BibitemOpen
  \bibfield  {author} {\bibinfo {author} {\bibfnamefont {Z.~C.}\ \bibnamefont
  {Zeng}}, \bibinfo {author} {\bibfnamefont {A.~H.~G.}\ \bibnamefont {Peter}},
  \bibinfo {author} {\bibfnamefont {X.}~\bibnamefont {Du}}, \bibinfo {author}
  {\bibfnamefont {S.}~\bibnamefont {Yang}}, \bibinfo {author} {\bibfnamefont
  {A.}~\bibnamefont {Benson}}, \bibinfo {author} {\bibfnamefont {F.-Y.}\
  \bibnamefont {Cyr-Racine}}, \bibinfo {author} {\bibfnamefont
  {F.}~\bibnamefont {Jiang}}, \bibinfo {author} {\bibfnamefont
  {C.}~\bibnamefont {Mace}}, \ and\ \bibinfo {author} {\bibfnamefont {R.~B.}\
  \bibnamefont {Metcalf}},\ }\href {\doibase 10.1103/PhysRevD.111.063001}
  {\bibfield  {journal} {\bibinfo  {journal} {Phys. Rev. D}\ }\textbf {\bibinfo
  {volume} {111}},\ \bibinfo {pages} {063001} (\bibinfo {year} {2025})},\
  \Eprint {http://arxiv.org/abs/2310.09910} {arXiv:2310.09910 [astro-ph.GA]}
  \BibitemShut {NoStop}%
\bibitem [{\citenamefont {Yang}\ \emph
  {et~al.}(2024{\natexlab{a}})\citenamefont {Yang}, \citenamefont {Nadler},
  \citenamefont {Yu},\ and\ \citenamefont {Zhong}}]{Yang:2023jwn}%
  \BibitemOpen
  \bibfield  {author} {\bibinfo {author} {\bibfnamefont {D.}~\bibnamefont
  {Yang}}, \bibinfo {author} {\bibfnamefont {E.~O.}\ \bibnamefont {Nadler}},
  \bibinfo {author} {\bibfnamefont {H.-B.}\ \bibnamefont {Yu}}, \ and\ \bibinfo
  {author} {\bibfnamefont {Y.-M.}\ \bibnamefont {Zhong}},\ }\href {\doibase
  10.1088/1475-7516/2024/02/032} {\bibfield  {journal} {\bibinfo  {journal}
  {JCAP}\ }\textbf {\bibinfo {volume} {02}},\ \bibinfo {pages} {032} (\bibinfo
  {year} {2024}{\natexlab{a}})},\ \Eprint {http://arxiv.org/abs/2305.16176}
  {arXiv:2305.16176 [astro-ph.CO]} \BibitemShut {NoStop}%
\bibitem [{\citenamefont {de~Blok}(2010)}]{deBlok:2009sp}%
  \BibitemOpen
  \bibfield  {author} {\bibinfo {author} {\bibfnamefont {W.~J.~G.}\
  \bibnamefont {de~Blok}},\ }\href {\doibase 10.1155/2010/789293} {\bibfield
  {journal} {\bibinfo  {journal} {Adv. Astron.}\ }\textbf {\bibinfo {volume}
  {2010}},\ \bibinfo {pages} {789293} (\bibinfo {year} {2010})},\ \Eprint
  {http://arxiv.org/abs/0910.3538} {arXiv:0910.3538 [astro-ph.CO]} \BibitemShut
  {NoStop}%
\bibitem [{\citenamefont {Boylan-Kolchin}\ \emph {et~al.}(2011)\citenamefont
  {Boylan-Kolchin}, \citenamefont {Bullock},\ and\ \citenamefont
  {Kaplinghat}}]{Boylan-Kolchin:2011qkt}%
  \BibitemOpen
  \bibfield  {author} {\bibinfo {author} {\bibfnamefont {M.}~\bibnamefont
  {Boylan-Kolchin}}, \bibinfo {author} {\bibfnamefont {J.~S.}\ \bibnamefont
  {Bullock}}, \ and\ \bibinfo {author} {\bibfnamefont {M.}~\bibnamefont
  {Kaplinghat}},\ }\href {\doibase 10.1111/j.1745-3933.2011.01074.x} {\bibfield
   {journal} {\bibinfo  {journal} {Mon. Not. Roy. Astron. Soc.}\ }\textbf
  {\bibinfo {volume} {415}},\ \bibinfo {pages} {L40} (\bibinfo {year}
  {2011})},\ \Eprint {http://arxiv.org/abs/1103.0007} {arXiv:1103.0007
  [astro-ph.CO]} \BibitemShut {NoStop}%
\bibitem [{\citenamefont {Gilman}\ \emph {et~al.}(2021)\citenamefont {Gilman},
  \citenamefont {Bovy}, \citenamefont {Treu}, \citenamefont {Nierenberg},
  \citenamefont {Birrer}, \citenamefont {Benson},\ and\ \citenamefont
  {Sameie}}]{Gilman:2021sdr}%
  \BibitemOpen
  \bibfield  {author} {\bibinfo {author} {\bibfnamefont {D.}~\bibnamefont
  {Gilman}}, \bibinfo {author} {\bibfnamefont {J.}~\bibnamefont {Bovy}},
  \bibinfo {author} {\bibfnamefont {T.}~\bibnamefont {Treu}}, \bibinfo {author}
  {\bibfnamefont {A.}~\bibnamefont {Nierenberg}}, \bibinfo {author}
  {\bibfnamefont {S.}~\bibnamefont {Birrer}}, \bibinfo {author} {\bibfnamefont
  {A.}~\bibnamefont {Benson}}, \ and\ \bibinfo {author} {\bibfnamefont
  {O.}~\bibnamefont {Sameie}},\ }\href {\doibase 10.1093/mnras/stab2335}
  {\bibfield  {journal} {\bibinfo  {journal} {Mon. Not. Roy. Astron. Soc.}\
  }\textbf {\bibinfo {volume} {507}},\ \bibinfo {pages} {2432} (\bibinfo {year}
  {2021})},\ \Eprint {http://arxiv.org/abs/2105.05259} {arXiv:2105.05259
  [astro-ph.CO]} \BibitemShut {NoStop}%
\bibitem [{\citenamefont {Gilman}\ \emph {et~al.}(2023)\citenamefont {Gilman},
  \citenamefont {Zhong},\ and\ \citenamefont {Bovy}}]{Gilman:2022ida}%
  \BibitemOpen
  \bibfield  {author} {\bibinfo {author} {\bibfnamefont {D.}~\bibnamefont
  {Gilman}}, \bibinfo {author} {\bibfnamefont {Y.-M.}\ \bibnamefont {Zhong}}, \
  and\ \bibinfo {author} {\bibfnamefont {J.}~\bibnamefont {Bovy}},\ }\href
  {\doibase 10.1103/PhysRevD.107.103008} {\bibfield  {journal} {\bibinfo
  {journal} {Phys. Rev. D}\ }\textbf {\bibinfo {volume} {107}},\ \bibinfo
  {pages} {103008} (\bibinfo {year} {2023})},\ \Eprint
  {http://arxiv.org/abs/2207.13111} {arXiv:2207.13111 [astro-ph.CO]}
  \BibitemShut {NoStop}%
\bibitem [{\citenamefont {Yu}(2025)}]{Yu:2025tmp}%
  \BibitemOpen
  \bibfield  {author} {\bibinfo {author} {\bibfnamefont {H.-B.}\ \bibnamefont
  {Yu}},\ }\href@noop {} {\  (\bibinfo {year} {2025})},\ \Eprint
  {http://arxiv.org/abs/2510.11006} {arXiv:2510.11006 [astro-ph.GA]}
  \BibitemShut {NoStop}%
\bibitem [{\citenamefont {Jiang}\ \emph {et~al.}(2026)\citenamefont {Jiang},
  \citenamefont {Jia}, \citenamefont {Zheng}, \citenamefont {Ho}, \citenamefont
  {Inayoshi}, \citenamefont {Shen}, \citenamefont {Vogelsberger},\ and\
  \citenamefont {Feng}}]{Jiang:2025jtr}%
  \BibitemOpen
  \bibfield  {author} {\bibinfo {author} {\bibfnamefont {F.}~\bibnamefont
  {Jiang}}, \bibinfo {author} {\bibfnamefont {Z.}~\bibnamefont {Jia}}, \bibinfo
  {author} {\bibfnamefont {H.}~\bibnamefont {Zheng}}, \bibinfo {author}
  {\bibfnamefont {L.~C.}\ \bibnamefont {Ho}}, \bibinfo {author} {\bibfnamefont
  {K.}~\bibnamefont {Inayoshi}}, \bibinfo {author} {\bibfnamefont
  {X.}~\bibnamefont {Shen}}, \bibinfo {author} {\bibfnamefont {M.}~\bibnamefont
  {Vogelsberger}}, \ and\ \bibinfo {author} {\bibfnamefont {W.-X.}\
  \bibnamefont {Feng}},\ }\href {\doibase 10.3847/2041-8213/ae247a} {\bibfield
  {journal} {\bibinfo  {journal} {Astrophys. J. Lett.}\ }\textbf {\bibinfo
  {volume} {996}},\ \bibinfo {pages} {L19} (\bibinfo {year} {2026})},\ \Eprint
  {http://arxiv.org/abs/2503.23710} {arXiv:2503.23710 [astro-ph.GA]}
  \BibitemShut {NoStop}%
\bibitem [{\citenamefont {Kong}\ \emph {et~al.}(2025)\citenamefont {Kong},
  \citenamefont {Nadler},\ and\ \citenamefont {Yu}}]{Kong:2025sqx}%
  \BibitemOpen
  \bibfield  {author} {\bibinfo {author} {\bibfnamefont {D.}~\bibnamefont
  {Kong}}, \bibinfo {author} {\bibfnamefont {E.~O.}\ \bibnamefont {Nadler}}, \
  and\ \bibinfo {author} {\bibfnamefont {H.-B.}\ \bibnamefont {Yu}},\
  }\href@noop {} {\  (\bibinfo {year} {2025})},\ \Eprint
  {http://arxiv.org/abs/2510.01491} {arXiv:2510.01491 [astro-ph.CO]}
  \BibitemShut {NoStop}%
\bibitem [{\citenamefont {Hou}\ \emph {et~al.}(2025)\citenamefont {Hou},
  \citenamefont {Yang}, \citenamefont {Li},\ and\ \citenamefont
  {Li}}]{Hou:2025gmv}%
  \BibitemOpen
  \bibfield  {author} {\bibinfo {author} {\bibfnamefont {S.}~\bibnamefont
  {Hou}}, \bibinfo {author} {\bibfnamefont {D.}~\bibnamefont {Yang}}, \bibinfo
  {author} {\bibfnamefont {N.}~\bibnamefont {Li}}, \ and\ \bibinfo {author}
  {\bibfnamefont {G.}~\bibnamefont {Li}},\ }\href {\doibase
  10.1088/1475-7516/2025/08/048} {\bibfield  {journal} {\bibinfo  {journal}
  {JCAP}\ }\textbf {\bibinfo {volume} {08}},\ \bibinfo {pages} {048} (\bibinfo
  {year} {2025})},\ \Eprint {http://arxiv.org/abs/2502.14964} {arXiv:2502.14964
  [astro-ph.CO]} \BibitemShut {NoStop}%
\bibitem [{\citenamefont {Shin}\ \emph {et~al.}(2021)\citenamefont {Shin} \emph
  {et~al.}}]{DES:2021qzb}%
  \BibitemOpen
  \bibfield  {author} {\bibinfo {author} {\bibfnamefont {T.}~\bibnamefont
  {Shin}} \emph {et~al.} (\bibinfo {collaboration} {DES}),\ }\href {\doibase
  10.1093/mnras/stab2505} {\bibfield  {journal} {\bibinfo  {journal} {Mon. Not.
  Roy. Astron. Soc.}\ }\textbf {\bibinfo {volume} {507}},\ \bibinfo {pages}
  {5758} (\bibinfo {year} {2021})},\ \Eprint {http://arxiv.org/abs/2105.05914}
  {arXiv:2105.05914 [astro-ph.CO]} \BibitemShut {NoStop}%
\bibitem [{\citenamefont {Adhikari}\ \emph {et~al.}(2024)\citenamefont
  {Adhikari}, \citenamefont {Banerjee}, \citenamefont {Jain}, \citenamefont
  {Hyeon-Shin},\ and\ \citenamefont {Zhong}}]{Adhikari:2024aff}%
  \BibitemOpen
  \bibfield  {author} {\bibinfo {author} {\bibfnamefont {S.}~\bibnamefont
  {Adhikari}}, \bibinfo {author} {\bibfnamefont {A.}~\bibnamefont {Banerjee}},
  \bibinfo {author} {\bibfnamefont {B.}~\bibnamefont {Jain}}, \bibinfo {author}
  {\bibfnamefont {T.}~\bibnamefont {Hyeon-Shin}}, \ and\ \bibinfo {author}
  {\bibfnamefont {Y.-M.}\ \bibnamefont {Zhong}},\ }\href@noop {} {\  (\bibinfo
  {year} {2024})},\ \Eprint {http://arxiv.org/abs/2401.05788} {arXiv:2401.05788
  [astro-ph.CO]} \BibitemShut {NoStop}%
\bibitem [{\citenamefont {Zhang}(2023)}]{Zhang:2022uzl}%
  \BibitemOpen
  \bibfield  {author} {\bibinfo {author} {\bibfnamefont {B.}~\bibnamefont
  {Zhang}},\ }\href {\doibase 10.1103/RevModPhys.95.035005} {\bibfield
  {journal} {\bibinfo  {journal} {Rev. Mod. Phys.}\ }\textbf {\bibinfo {volume}
  {95}},\ \bibinfo {pages} {035005} (\bibinfo {year} {2023})},\ \Eprint
  {http://arxiv.org/abs/2212.03972} {arXiv:2212.03972 [astro-ph.HE]}
  \BibitemShut {NoStop}%
\bibitem [{\citenamefont {Petroff}\ \emph {et~al.}(2019)\citenamefont
  {Petroff}, \citenamefont {Hessels},\ and\ \citenamefont
  {Lorimer}}]{Petroff:2019tty}%
  \BibitemOpen
  \bibfield  {author} {\bibinfo {author} {\bibfnamefont {E.}~\bibnamefont
  {Petroff}}, \bibinfo {author} {\bibfnamefont {J.~W.~T.}\ \bibnamefont
  {Hessels}}, \ and\ \bibinfo {author} {\bibfnamefont {D.~R.}\ \bibnamefont
  {Lorimer}},\ }\href {\doibase 10.1007/s00159-019-0116-6} {\bibfield
  {journal} {\bibinfo  {journal} {Astron. Astrophys. Rev.}\ }\textbf {\bibinfo
  {volume} {27}},\ \bibinfo {pages} {4} (\bibinfo {year} {2019})},\ \Eprint
  {http://arxiv.org/abs/1904.07947} {arXiv:1904.07947 [astro-ph.HE]}
  \BibitemShut {NoStop}%
\bibitem [{\citenamefont {{Chatterjee}}\ \emph {et~al.}(2017)\citenamefont
  {{Chatterjee}}, \citenamefont {{Law}}, \citenamefont {{Wharton}},
  \citenamefont {{Burke-Spolaor}}, \citenamefont {{Hessels}}, \citenamefont
  {{Bower}}, \citenamefont {{Cordes}}, \citenamefont {{Tendulkar}},
  \citenamefont {{Bassa}}, \citenamefont {{Demorest}}, \citenamefont
  {{Butler}}, \citenamefont {{Seymour}}, \citenamefont {{Scholz}},
  \citenamefont {{Abruzzo}}, \citenamefont {{Bogdanov}}, \citenamefont
  {{Kaspi}}, \citenamefont {{Keimpema}}, \citenamefont {{Lazio}}, \citenamefont
  {{Marcote}}, \citenamefont {{McLaughlin}}, \citenamefont {{Paragi}},
  \citenamefont {{Ransom}}, \citenamefont {{Rupen}}, \citenamefont
  {{Spitler}},\ and\ \citenamefont {{van Langevelde}}}]{2017Natur.541...58C}%
  \BibitemOpen
  \bibfield  {author} {\bibinfo {author} {\bibfnamefont {S.}~\bibnamefont
  {{Chatterjee}}}, \bibinfo {author} {\bibfnamefont {C.~J.}\ \bibnamefont
  {{Law}}}, \bibinfo {author} {\bibfnamefont {R.~S.}\ \bibnamefont
  {{Wharton}}}, \bibinfo {author} {\bibfnamefont {S.}~\bibnamefont
  {{Burke-Spolaor}}}, \bibinfo {author} {\bibfnamefont {J.~W.~T.}\ \bibnamefont
  {{Hessels}}}, \bibinfo {author} {\bibfnamefont {G.~C.}\ \bibnamefont
  {{Bower}}}, \bibinfo {author} {\bibfnamefont {J.~M.}\ \bibnamefont
  {{Cordes}}}, \bibinfo {author} {\bibfnamefont {S.~P.}\ \bibnamefont
  {{Tendulkar}}}, \bibinfo {author} {\bibfnamefont {C.~G.}\ \bibnamefont
  {{Bassa}}}, \bibinfo {author} {\bibfnamefont {P.}~\bibnamefont {{Demorest}}},
  \bibinfo {author} {\bibfnamefont {B.~J.}\ \bibnamefont {{Butler}}}, \bibinfo
  {author} {\bibfnamefont {A.}~\bibnamefont {{Seymour}}}, \bibinfo {author}
  {\bibfnamefont {P.}~\bibnamefont {{Scholz}}}, \bibinfo {author}
  {\bibfnamefont {M.~W.}\ \bibnamefont {{Abruzzo}}}, \bibinfo {author}
  {\bibfnamefont {S.}~\bibnamefont {{Bogdanov}}}, \bibinfo {author}
  {\bibfnamefont {V.~M.}\ \bibnamefont {{Kaspi}}}, \bibinfo {author}
  {\bibfnamefont {A.}~\bibnamefont {{Keimpema}}}, \bibinfo {author}
  {\bibfnamefont {T.~J.~W.}\ \bibnamefont {{Lazio}}}, \bibinfo {author}
  {\bibfnamefont {B.}~\bibnamefont {{Marcote}}}, \bibinfo {author}
  {\bibfnamefont {M.~A.}\ \bibnamefont {{McLaughlin}}}, \bibinfo {author}
  {\bibfnamefont {Z.}~\bibnamefont {{Paragi}}}, \bibinfo {author}
  {\bibfnamefont {S.~M.}\ \bibnamefont {{Ransom}}}, \bibinfo {author}
  {\bibfnamefont {M.}~\bibnamefont {{Rupen}}}, \bibinfo {author} {\bibfnamefont
  {L.~G.}\ \bibnamefont {{Spitler}}}, \ and\ \bibinfo {author} {\bibfnamefont
  {H.~J.}\ \bibnamefont {{van Langevelde}}},\ }\href {\doibase
  10.1038/nature20797} {\bibfield  {journal} {\bibinfo  {journal} {\nat}\
  }\textbf {\bibinfo {volume} {541}},\ \bibinfo {pages} {58} (\bibinfo {year}
  {2017})},\ \Eprint {http://arxiv.org/abs/1701.01098} {arXiv:1701.01098
  [astro-ph.HE]} \BibitemShut {NoStop}%
\bibitem [{\citenamefont {{Bannister}}\ \emph {et~al.}(2019)\citenamefont
  {{Bannister}}, \citenamefont {{Deller}}, \citenamefont {{Phillips}},
  \citenamefont {{Macquart}}, \citenamefont {{Prochaska}}, \citenamefont
  {{Tejos}}, \citenamefont {{Ryder}}, \citenamefont {{Sadler}}, \citenamefont
  {{Shannon}}, \citenamefont {{Simha}}, \citenamefont {{Day}}, \citenamefont
  {{McQuinn}}, \citenamefont {{North-Hickey}}, \citenamefont {{Bhandari}},
  \citenamefont {{Arcus}}, \citenamefont {{Bennert}}, \citenamefont
  {{Burchett}}, \citenamefont {{Bouwhuis}}, \citenamefont {{Dodson}},
  \citenamefont {{Ekers}}, \citenamefont {{Farah}}, \citenamefont {{Flynn}},
  \citenamefont {{James}}, \citenamefont {{Kerr}}, \citenamefont {{Lenc}},
  \citenamefont {{Mahony}}, \citenamefont {{O'Meara}}, \citenamefont
  {{Os{\l}owski}}, \citenamefont {{Qiu}}, \citenamefont {{Treu}}, \citenamefont
  {{U}}, \citenamefont {{Bateman}}, \citenamefont {{Bock}}, \citenamefont
  {{Bolton}}, \citenamefont {{Brown}}, \citenamefont {{Bunton}}, \citenamefont
  {{Chippendale}}, \citenamefont {{Cooray}}, \citenamefont {{Cornwell}},
  \citenamefont {{Gupta}}, \citenamefont {{Hayman}}, \citenamefont
  {{Kesteven}}, \citenamefont {{Koribalski}}, \citenamefont {{MacLeod}},
  \citenamefont {{McClure-Griffiths}}, \citenamefont {{Neuhold}}, \citenamefont
  {{Norris}}, \citenamefont {{Pilawa}}, \citenamefont {{Qiao}}, \citenamefont
  {{Reynolds}}, \citenamefont {{Roxby}}, \citenamefont {{Shimwell}},
  \citenamefont {{Voronkov}},\ and\ \citenamefont
  {{Wilson}}}]{2019Sci...365..565B}%
  \BibitemOpen
  \bibfield  {author} {\bibinfo {author} {\bibfnamefont {K.~W.}\ \bibnamefont
  {{Bannister}}}, \bibinfo {author} {\bibfnamefont {A.~T.}\ \bibnamefont
  {{Deller}}}, \bibinfo {author} {\bibfnamefont {C.}~\bibnamefont
  {{Phillips}}}, \bibinfo {author} {\bibfnamefont {J.-P.}\ \bibnamefont
  {{Macquart}}}, \bibinfo {author} {\bibfnamefont {J.~X.}\ \bibnamefont
  {{Prochaska}}}, \bibinfo {author} {\bibfnamefont {N.}~\bibnamefont
  {{Tejos}}}, \bibinfo {author} {\bibfnamefont {S.~D.}\ \bibnamefont
  {{Ryder}}}, \bibinfo {author} {\bibfnamefont {E.~M.}\ \bibnamefont
  {{Sadler}}}, \bibinfo {author} {\bibfnamefont {R.~M.}\ \bibnamefont
  {{Shannon}}}, \bibinfo {author} {\bibfnamefont {S.}~\bibnamefont {{Simha}}},
  \bibinfo {author} {\bibfnamefont {C.~K.}\ \bibnamefont {{Day}}}, \bibinfo
  {author} {\bibfnamefont {M.}~\bibnamefont {{McQuinn}}}, \bibinfo {author}
  {\bibfnamefont {F.~O.}\ \bibnamefont {{North-Hickey}}}, \bibinfo {author}
  {\bibfnamefont {S.}~\bibnamefont {{Bhandari}}}, \bibinfo {author}
  {\bibfnamefont {W.~R.}\ \bibnamefont {{Arcus}}}, \bibinfo {author}
  {\bibfnamefont {V.~N.}\ \bibnamefont {{Bennert}}}, \bibinfo {author}
  {\bibfnamefont {J.}~\bibnamefont {{Burchett}}}, \bibinfo {author}
  {\bibfnamefont {M.}~\bibnamefont {{Bouwhuis}}}, \bibinfo {author}
  {\bibfnamefont {R.}~\bibnamefont {{Dodson}}}, \bibinfo {author}
  {\bibfnamefont {R.~D.}\ \bibnamefont {{Ekers}}}, \bibinfo {author}
  {\bibfnamefont {W.}~\bibnamefont {{Farah}}}, \bibinfo {author} {\bibfnamefont
  {C.}~\bibnamefont {{Flynn}}}, \bibinfo {author} {\bibfnamefont {C.~W.}\
  \bibnamefont {{James}}}, \bibinfo {author} {\bibfnamefont {M.}~\bibnamefont
  {{Kerr}}}, \bibinfo {author} {\bibfnamefont {E.}~\bibnamefont {{Lenc}}},
  \bibinfo {author} {\bibfnamefont {E.~K.}\ \bibnamefont {{Mahony}}}, \bibinfo
  {author} {\bibfnamefont {J.}~\bibnamefont {{O'Meara}}}, \bibinfo {author}
  {\bibfnamefont {S.}~\bibnamefont {{Os{\l}owski}}}, \bibinfo {author}
  {\bibfnamefont {H.}~\bibnamefont {{Qiu}}}, \bibinfo {author} {\bibfnamefont
  {T.}~\bibnamefont {{Treu}}}, \bibinfo {author} {\bibfnamefont
  {V.}~\bibnamefont {{U}}}, \bibinfo {author} {\bibfnamefont {T.~J.}\
  \bibnamefont {{Bateman}}}, \bibinfo {author} {\bibfnamefont {D.~C.-J.}\
  \bibnamefont {{Bock}}}, \bibinfo {author} {\bibfnamefont {R.~J.}\
  \bibnamefont {{Bolton}}}, \bibinfo {author} {\bibfnamefont {A.}~\bibnamefont
  {{Brown}}}, \bibinfo {author} {\bibfnamefont {J.~D.}\ \bibnamefont
  {{Bunton}}}, \bibinfo {author} {\bibfnamefont {A.~P.}\ \bibnamefont
  {{Chippendale}}}, \bibinfo {author} {\bibfnamefont {F.~R.}\ \bibnamefont
  {{Cooray}}}, \bibinfo {author} {\bibfnamefont {T.}~\bibnamefont
  {{Cornwell}}}, \bibinfo {author} {\bibfnamefont {N.}~\bibnamefont {{Gupta}}},
  \bibinfo {author} {\bibfnamefont {D.~B.}\ \bibnamefont {{Hayman}}}, \bibinfo
  {author} {\bibfnamefont {M.}~\bibnamefont {{Kesteven}}}, \bibinfo {author}
  {\bibfnamefont {B.~S.}\ \bibnamefont {{Koribalski}}}, \bibinfo {author}
  {\bibfnamefont {A.}~\bibnamefont {{MacLeod}}}, \bibinfo {author}
  {\bibfnamefont {N.~M.}\ \bibnamefont {{McClure-Griffiths}}}, \bibinfo
  {author} {\bibfnamefont {S.}~\bibnamefont {{Neuhold}}}, \bibinfo {author}
  {\bibfnamefont {R.~P.}\ \bibnamefont {{Norris}}}, \bibinfo {author}
  {\bibfnamefont {M.~A.}\ \bibnamefont {{Pilawa}}}, \bibinfo {author}
  {\bibfnamefont {R.-Y.}\ \bibnamefont {{Qiao}}}, \bibinfo {author}
  {\bibfnamefont {J.}~\bibnamefont {{Reynolds}}}, \bibinfo {author}
  {\bibfnamefont {D.~N.}\ \bibnamefont {{Roxby}}}, \bibinfo {author}
  {\bibfnamefont {T.~W.}\ \bibnamefont {{Shimwell}}}, \bibinfo {author}
  {\bibfnamefont {M.~A.}\ \bibnamefont {{Voronkov}}}, \ and\ \bibinfo {author}
  {\bibfnamefont {C.~D.}\ \bibnamefont {{Wilson}}},\ }\href {\doibase
  10.1126/science.aaw5903} {\bibfield  {journal} {\bibinfo  {journal}
  {Science}\ }\textbf {\bibinfo {volume} {365}},\ \bibinfo {pages} {565}
  (\bibinfo {year} {2019})},\ \Eprint {http://arxiv.org/abs/1906.11476}
  {arXiv:1906.11476 [astro-ph.HE]} \BibitemShut {NoStop}%
\bibitem [{\citenamefont {{Law}}\ \emph {et~al.}(2024)\citenamefont {{Law}},
  \citenamefont {{Sharma}}, \citenamefont {{Ravi}}, \citenamefont {{Chen}},
  \citenamefont {{Catha}}, \citenamefont {{Connor}}, \citenamefont {{Faber}},
  \citenamefont {{Hallinan}}, \citenamefont {{Harnach}}, \citenamefont
  {{Hellbourg}}, \citenamefont {{Hobbs}}, \citenamefont {{Hodge}},
  \citenamefont {{Hodges}}, \citenamefont {{Lamb}}, \citenamefont
  {{Rasmussen}}, \citenamefont {{Sherman}}, \citenamefont {{Shi}},
  \citenamefont {{Simard}}, \citenamefont {{Squillace}}, \citenamefont
  {{Weinreb}}, \citenamefont {{Woody}},\ and\ \citenamefont
  {{Yurk}}}]{2024ApJ...967...29L}%
  \BibitemOpen
  \bibfield  {author} {\bibinfo {author} {\bibfnamefont {C.~J.}\ \bibnamefont
  {{Law}}}, \bibinfo {author} {\bibfnamefont {K.}~\bibnamefont {{Sharma}}},
  \bibinfo {author} {\bibfnamefont {V.}~\bibnamefont {{Ravi}}}, \bibinfo
  {author} {\bibfnamefont {G.}~\bibnamefont {{Chen}}}, \bibinfo {author}
  {\bibfnamefont {M.}~\bibnamefont {{Catha}}}, \bibinfo {author} {\bibfnamefont
  {L.}~\bibnamefont {{Connor}}}, \bibinfo {author} {\bibfnamefont {J.~T.}\
  \bibnamefont {{Faber}}}, \bibinfo {author} {\bibfnamefont {G.}~\bibnamefont
  {{Hallinan}}}, \bibinfo {author} {\bibfnamefont {C.}~\bibnamefont
  {{Harnach}}}, \bibinfo {author} {\bibfnamefont {G.}~\bibnamefont
  {{Hellbourg}}}, \bibinfo {author} {\bibfnamefont {R.}~\bibnamefont
  {{Hobbs}}}, \bibinfo {author} {\bibfnamefont {D.}~\bibnamefont {{Hodge}}},
  \bibinfo {author} {\bibfnamefont {M.}~\bibnamefont {{Hodges}}}, \bibinfo
  {author} {\bibfnamefont {J.~W.}\ \bibnamefont {{Lamb}}}, \bibinfo {author}
  {\bibfnamefont {P.}~\bibnamefont {{Rasmussen}}}, \bibinfo {author}
  {\bibfnamefont {M.~B.}\ \bibnamefont {{Sherman}}}, \bibinfo {author}
  {\bibfnamefont {J.}~\bibnamefont {{Shi}}}, \bibinfo {author} {\bibfnamefont
  {D.}~\bibnamefont {{Simard}}}, \bibinfo {author} {\bibfnamefont
  {R.}~\bibnamefont {{Squillace}}}, \bibinfo {author} {\bibfnamefont
  {S.}~\bibnamefont {{Weinreb}}}, \bibinfo {author} {\bibfnamefont {D.~P.}\
  \bibnamefont {{Woody}}}, \ and\ \bibinfo {author} {\bibfnamefont {N.~Y.}\
  \bibnamefont {{Yurk}}},\ }\href {\doibase 10.3847/1538-4357/ad3736}
  {\bibfield  {journal} {\bibinfo  {journal} {\apj}\ }\textbf {\bibinfo
  {volume} {967}},\ \bibinfo {eid} {29} (\bibinfo {year} {2024})},\ \Eprint
  {http://arxiv.org/abs/2307.03344} {arXiv:2307.03344 [astro-ph.HE]}
  \BibitemShut {NoStop}%
\bibitem [{\citenamefont {{Zhang}}(2023)}]{2023RvMP...95c5005Z}%
  \BibitemOpen
  \bibfield  {author} {\bibinfo {author} {\bibfnamefont {B.}~\bibnamefont
  {{Zhang}}},\ }\href {\doibase 10.1103/RevModPhys.95.035005} {\bibfield
  {journal} {\bibinfo  {journal} {Reviews of Modern Physics}\ }\textbf
  {\bibinfo {volume} {95}},\ \bibinfo {eid} {035005} (\bibinfo {year}
  {2023})},\ \Eprint {http://arxiv.org/abs/2212.03972} {arXiv:2212.03972
  [astro-ph.HE]} \BibitemShut {NoStop}%
\bibitem [{\citenamefont {Mu{\~n}oz}\ \emph {et~al.}(2016)\citenamefont
  {Mu{\~n}oz}, \citenamefont {Kovetz}, \citenamefont {Dai},\ and\ \citenamefont
  {Kamionkowski}}]{Munoz:2016tmg}%
  \BibitemOpen
  \bibfield  {author} {\bibinfo {author} {\bibfnamefont {J.~B.}\ \bibnamefont
  {Mu{\~n}oz}}, \bibinfo {author} {\bibfnamefont {E.~D.}\ \bibnamefont
  {Kovetz}}, \bibinfo {author} {\bibfnamefont {L.}~\bibnamefont {Dai}}, \ and\
  \bibinfo {author} {\bibfnamefont {M.}~\bibnamefont {Kamionkowski}},\ }\href
  {\doibase 10.1103/PhysRevLett.117.091301} {\bibfield  {journal} {\bibinfo
  {journal} {Phys. Rev. Lett.}\ }\textbf {\bibinfo {volume} {117}},\ \bibinfo
  {pages} {091301} (\bibinfo {year} {2016})},\ \Eprint
  {http://arxiv.org/abs/1605.00008} {arXiv:1605.00008 [astro-ph.CO]}
  \BibitemShut {NoStop}%
\bibitem [{\citenamefont {Li}\ \emph {et~al.}(2018)\citenamefont {Li},
  \citenamefont {Gao}, \citenamefont {Ding}, \citenamefont {Wang},\ and\
  \citenamefont {Zhang}}]{Li:2017mek}%
  \BibitemOpen
  \bibfield  {author} {\bibinfo {author} {\bibfnamefont {Z.-X.}\ \bibnamefont
  {Li}}, \bibinfo {author} {\bibfnamefont {H.}~\bibnamefont {Gao}}, \bibinfo
  {author} {\bibfnamefont {X.-H.}\ \bibnamefont {Ding}}, \bibinfo {author}
  {\bibfnamefont {G.-J.}\ \bibnamefont {Wang}}, \ and\ \bibinfo {author}
  {\bibfnamefont {B.}~\bibnamefont {Zhang}},\ }\href {\doibase
  10.1038/s41467-018-06303-0} {\bibfield  {journal} {\bibinfo  {journal}
  {Nature Commun.}\ }\textbf {\bibinfo {volume} {9}},\ \bibinfo {pages} {3833}
  (\bibinfo {year} {2018})},\ \Eprint {http://arxiv.org/abs/1708.06357}
  {arXiv:1708.06357 [astro-ph.CO]} \BibitemShut {NoStop}%
\bibitem [{\citenamefont {Wang}\ and\ \citenamefont
  {Wang}(2018)}]{Wang:2018ydd}%
  \BibitemOpen
  \bibfield  {author} {\bibinfo {author} {\bibfnamefont {Y.~K.}\ \bibnamefont
  {Wang}}\ and\ \bibinfo {author} {\bibfnamefont {F.~Y.}\ \bibnamefont
  {Wang}},\ }\href {\doibase 10.1051/0004-6361/201731160} {\bibfield  {journal}
  {\bibinfo  {journal} {Astron. Astrophys.}\ }\textbf {\bibinfo {volume}
  {614}},\ \bibinfo {pages} {A50} (\bibinfo {year} {2018})},\ \Eprint
  {http://arxiv.org/abs/1801.07360} {arXiv:1801.07360 [astro-ph.CO]}
  \BibitemShut {NoStop}%
\bibitem [{\citenamefont {Wucknitz}\ \emph {et~al.}(2021)\citenamefont
  {Wucknitz}, \citenamefont {Spitler},\ and\ \citenamefont
  {Pen}}]{Wucknitz:2020spz}%
  \BibitemOpen
  \bibfield  {author} {\bibinfo {author} {\bibfnamefont {O.}~\bibnamefont
  {Wucknitz}}, \bibinfo {author} {\bibfnamefont {L.~G.}\ \bibnamefont
  {Spitler}}, \ and\ \bibinfo {author} {\bibfnamefont {U.~L.}\ \bibnamefont
  {Pen}},\ }\href {\doibase 10.1051/0004-6361/202038248} {\bibfield  {journal}
  {\bibinfo  {journal} {Astron. Astrophys.}\ }\textbf {\bibinfo {volume}
  {645}},\ \bibinfo {pages} {A44} (\bibinfo {year} {2021})},\ \Eprint
  {http://arxiv.org/abs/2004.11643} {arXiv:2004.11643 [astro-ph.CO]}
  \BibitemShut {NoStop}%
\bibitem [{\citenamefont {Adi}\ and\ \citenamefont
  {Kovetz}(2021)}]{Adi:2021uuw}%
  \BibitemOpen
  \bibfield  {author} {\bibinfo {author} {\bibfnamefont {T.}~\bibnamefont
  {Adi}}\ and\ \bibinfo {author} {\bibfnamefont {E.~D.}\ \bibnamefont
  {Kovetz}},\ }\href {\doibase 10.1103/PhysRevD.104.103515} {\bibfield
  {journal} {\bibinfo  {journal} {Phys. Rev. D}\ }\textbf {\bibinfo {volume}
  {104}},\ \bibinfo {pages} {103515} (\bibinfo {year} {2021})},\ \Eprint
  {http://arxiv.org/abs/2109.00403} {arXiv:2109.00403 [astro-ph.CO]}
  \BibitemShut {NoStop}%
\bibitem [{\citenamefont {Leung}\ \emph {et~al.}(2022)\citenamefont {Leung}
  \emph {et~al.}}]{Leung:2022vcx}%
  \BibitemOpen
  \bibfield  {author} {\bibinfo {author} {\bibfnamefont {C.}~\bibnamefont
  {Leung}} \emph {et~al.},\ }\href {\doibase 10.1103/PhysRevD.106.043017}
  {\bibfield  {journal} {\bibinfo  {journal} {Phys. Rev. D}\ }\textbf {\bibinfo
  {volume} {106}},\ \bibinfo {pages} {043017} (\bibinfo {year} {2022})},\
  \Eprint {http://arxiv.org/abs/2204.06001} {arXiv:2204.06001 [astro-ph.HE]}
  \BibitemShut {NoStop}%
\bibitem [{\citenamefont {Singh}\ \emph {et~al.}(2023)\citenamefont {Singh},
  \citenamefont {Kapadia}, \citenamefont {Basak}, \citenamefont {Ajith},\ and\
  \citenamefont {Tendulkar}}]{Singh:2023hbd}%
  \BibitemOpen
  \bibfield  {author} {\bibinfo {author} {\bibfnamefont {M.~K.}\ \bibnamefont
  {Singh}}, \bibinfo {author} {\bibfnamefont {S.~J.}\ \bibnamefont {Kapadia}},
  \bibinfo {author} {\bibfnamefont {S.}~\bibnamefont {Basak}}, \bibinfo
  {author} {\bibfnamefont {P.}~\bibnamefont {Ajith}}, \ and\ \bibinfo {author}
  {\bibfnamefont {S.~P.}\ \bibnamefont {Tendulkar}},\ }\href {\doibase
  10.1093/mnras/stad3376} {\bibfield  {journal} {\bibinfo  {journal} {Mon. Not.
  Roy. Astron. Soc.}\ }\textbf {\bibinfo {volume} {527}},\ \bibinfo {pages}
  {4234} (\bibinfo {year} {2023})},\ \Eprint {http://arxiv.org/abs/2304.02879}
  {arXiv:2304.02879 [astro-ph.HE]} \BibitemShut {NoStop}%
\bibitem [{\citenamefont {Dall'Osso}\ \emph {et~al.}(2024)\citenamefont
  {Dall'Osso}, \citenamefont {La~Placa}, \citenamefont {Stella}, \citenamefont
  {Bakala},\ and\ \citenamefont {Possenti}}]{DallOsso:2024sjv}%
  \BibitemOpen
  \bibfield  {author} {\bibinfo {author} {\bibfnamefont {S.}~\bibnamefont
  {Dall'Osso}}, \bibinfo {author} {\bibfnamefont {R.}~\bibnamefont {La~Placa}},
  \bibinfo {author} {\bibfnamefont {L.}~\bibnamefont {Stella}}, \bibinfo
  {author} {\bibfnamefont {P.}~\bibnamefont {Bakala}}, \ and\ \bibinfo {author}
  {\bibfnamefont {A.}~\bibnamefont {Possenti}},\ }\href {\doibase
  10.3847/1538-4357/ad5f1c} {\bibfield  {journal} {\bibinfo  {journal}
  {Astrophys. J.}\ }\textbf {\bibinfo {volume} {973}},\ \bibinfo {pages} {123}
  (\bibinfo {year} {2024})},\ \Eprint {http://arxiv.org/abs/2407.04095}
  {arXiv:2407.04095 [astro-ph.HE]} \BibitemShut {NoStop}%
\bibitem [{\citenamefont {Pastor-Marazuela}(2025)}]{Pastor-Marazuela:2024lcd}%
  \BibitemOpen
  \bibfield  {author} {\bibinfo {author} {\bibfnamefont {I.}~\bibnamefont
  {Pastor-Marazuela}},\ }\href {\doibase 10.1098/rsta.2024.0121} {\bibfield
  {journal} {\bibinfo  {journal} {Philosophical Transactions of the Royal
  Society A: Mathematical, Physical and Engineering Sciences}\ }\textbf
  {\bibinfo {volume} {383}},\ \bibinfo {pages} {20240121} (\bibinfo {year}
  {2025})},\ \Eprint
  {http://arxiv.org/abs/https://royalsocietypublishing.org/rsta/article-pdf/doi/10.1098/rsta.2024.0121/2817908/rsta.2024.0121.pdf}
  {https://royalsocietypublishing.org/rsta/article-pdf/doi/10.1098/rsta.2024.0121/2817908/rsta.2024.0121.pdf}
  \BibitemShut {NoStop}%
\bibitem [{\citenamefont {Kader}\ \emph {et~al.}(2024)\citenamefont {Kader},
  \citenamefont {Dobbs}, \citenamefont {Leung}, \citenamefont {Masui},\ and\
  \citenamefont {Sammons}}]{Kader:2024uqm}%
  \BibitemOpen
  \bibfield  {author} {\bibinfo {author} {\bibfnamefont {Z.}~\bibnamefont
  {Kader}}, \bibinfo {author} {\bibfnamefont {M.}~\bibnamefont {Dobbs}},
  \bibinfo {author} {\bibfnamefont {C.}~\bibnamefont {Leung}}, \bibinfo
  {author} {\bibfnamefont {K.~W.}\ \bibnamefont {Masui}}, \ and\ \bibinfo
  {author} {\bibfnamefont {M.~W.}\ \bibnamefont {Sammons}},\ }\href {\doibase
  10.1103/PhysRevD.110.123027} {\bibfield  {journal} {\bibinfo  {journal}
  {Phys. Rev. D}\ }\textbf {\bibinfo {volume} {110}},\ \bibinfo {pages}
  {123027} (\bibinfo {year} {2024})},\ \Eprint
  {http://arxiv.org/abs/2407.04097} {arXiv:2407.04097 [astro-ph.HE]}
  \BibitemShut {NoStop}%
\bibitem [{\citenamefont {Xiao}\ \emph {et~al.}(2024)\citenamefont {Xiao},
  \citenamefont {Dai},\ and\ \citenamefont {McQuinn}}]{Xiao:2024qay}%
  \BibitemOpen
  \bibfield  {author} {\bibinfo {author} {\bibfnamefont {H.}~\bibnamefont
  {Xiao}}, \bibinfo {author} {\bibfnamefont {L.}~\bibnamefont {Dai}}, \ and\
  \bibinfo {author} {\bibfnamefont {M.}~\bibnamefont {McQuinn}},\ }\href
  {\doibase 10.1103/PhysRevD.110.023516} {\bibfield  {journal} {\bibinfo
  {journal} {Phys. Rev. D}\ }\textbf {\bibinfo {volume} {110}},\ \bibinfo
  {pages} {023516} (\bibinfo {year} {2024})},\ \Eprint
  {http://arxiv.org/abs/2401.08862} {arXiv:2401.08862 [astro-ph.CO]}
  \BibitemShut {NoStop}%
\bibitem [{\citenamefont {Xiong}\ \emph {et~al.}(2026)\citenamefont {Xiong},
  \citenamefont {Xiao}, \citenamefont {Jiang}, \citenamefont {Lai},
  \citenamefont {Zhang}, \citenamefont {Zhao},\ and\ \citenamefont
  {You}}]{Xiong:2025gtw}%
  \BibitemOpen
  \bibfield  {author} {\bibinfo {author} {\bibfnamefont {S.-W.}\ \bibnamefont
  {Xiong}}, \bibinfo {author} {\bibfnamefont {S.}~\bibnamefont {Xiao}},
  \bibinfo {author} {\bibfnamefont {Z.-H.}\ \bibnamefont {Jiang}}, \bibinfo
  {author} {\bibfnamefont {Y.}~\bibnamefont {Lai}}, \bibinfo {author}
  {\bibfnamefont {Y.-Q.}\ \bibnamefont {Zhang}}, \bibinfo {author}
  {\bibfnamefont {R.-S.}\ \bibnamefont {Zhao}}, \ and\ \bibinfo {author}
  {\bibfnamefont {Z.-Y.}\ \bibnamefont {You}},\ }\href {\doibase
  10.3847/1538-4357/ae22d3} {\bibfield  {journal} {\bibinfo  {journal}
  {Astrophys. J.}\ }\textbf {\bibinfo {volume} {996}},\ \bibinfo {pages} {29}
  (\bibinfo {year} {2026})}\BibitemShut {NoStop}%
\bibitem [{\citenamefont {Meena}\ and\ \citenamefont
  {Saha}(2025)}]{Meena:2025cdo}%
  \BibitemOpen
  \bibfield  {author} {\bibinfo {author} {\bibfnamefont {A.~K.}\ \bibnamefont
  {Meena}}\ and\ \bibinfo {author} {\bibfnamefont {P.}~\bibnamefont {Saha}},\
  }\href {\doibase 10.1103/vxts-hbz7} {\bibfield  {journal} {\bibinfo
  {journal} {Phys. Rev. D}\ }\textbf {\bibinfo {volume} {112}},\ \bibinfo
  {pages} {123012} (\bibinfo {year} {2025})},\ \Eprint
  {http://arxiv.org/abs/2507.20305} {arXiv:2507.20305 [astro-ph.GA]}
  \BibitemShut {NoStop}%
\bibitem [{\citenamefont {Li}\ \emph {et~al.}(2025)\citenamefont {Li},
  \citenamefont {Wang}, \citenamefont {Zhao},\ and\ \citenamefont
  {Li}}]{Li:2025wdc}%
  \BibitemOpen
  \bibfield  {author} {\bibinfo {author} {\bibfnamefont {J.-H.}\ \bibnamefont
  {Li}}, \bibinfo {author} {\bibfnamefont {S.-J.}\ \bibnamefont {Wang}},
  \bibinfo {author} {\bibfnamefont {X.-Y.}\ \bibnamefont {Zhao}}, \ and\
  \bibinfo {author} {\bibfnamefont {N.}~\bibnamefont {Li}},\ }\href {\doibase
  10.3390/universe11090311} {\bibfield  {journal} {\bibinfo  {journal}
  {Universe}\ }\textbf {\bibinfo {volume} {11}},\ \bibinfo {pages} {311}
  (\bibinfo {year} {2025})}\BibitemShut {NoStop}%
\bibitem [{\citenamefont {Ludlow}\ \emph {et~al.}(2016)\citenamefont {Ludlow},
  \citenamefont {Bose}, \citenamefont {Angulo}, \citenamefont {Wang},
  \citenamefont {Hellwing}, \citenamefont {Navarro}, \citenamefont {Cole},\
  and\ \citenamefont {Frenk}}]{Ludlow:2016ifl}%
  \BibitemOpen
  \bibfield  {author} {\bibinfo {author} {\bibfnamefont {A.~D.}\ \bibnamefont
  {Ludlow}}, \bibinfo {author} {\bibfnamefont {S.}~\bibnamefont {Bose}},
  \bibinfo {author} {\bibfnamefont {R.~E.}\ \bibnamefont {Angulo}}, \bibinfo
  {author} {\bibfnamefont {L.}~\bibnamefont {Wang}}, \bibinfo {author}
  {\bibfnamefont {W.~A.}\ \bibnamefont {Hellwing}}, \bibinfo {author}
  {\bibfnamefont {J.~F.}\ \bibnamefont {Navarro}}, \bibinfo {author}
  {\bibfnamefont {S.}~\bibnamefont {Cole}}, \ and\ \bibinfo {author}
  {\bibfnamefont {C.~S.}\ \bibnamefont {Frenk}},\ }\href {\doibase
  10.1093/mnras/stw1046} {\bibfield  {journal} {\bibinfo  {journal} {Mon. Not.
  Roy. Astron. Soc.}\ }\textbf {\bibinfo {volume} {460}},\ \bibinfo {pages}
  {1214} (\bibinfo {year} {2016})},\ \Eprint {http://arxiv.org/abs/1601.02624}
  {arXiv:1601.02624 [astro-ph.CO]} \BibitemShut {NoStop}%
\bibitem [{\citenamefont {Zeng}\ \emph {et~al.}(2022)\citenamefont {Zeng},
  \citenamefont {Peter}, \citenamefont {Du}, \citenamefont {Benson},
  \citenamefont {Kim}, \citenamefont {Jiang}, \citenamefont {Cyr-Racine},\ and\
  \citenamefont {Vogelsberger}}]{Zeng:2021ldo}%
  \BibitemOpen
  \bibfield  {author} {\bibinfo {author} {\bibfnamefont {Z.~C.}\ \bibnamefont
  {Zeng}}, \bibinfo {author} {\bibfnamefont {A.~H.~G.}\ \bibnamefont {Peter}},
  \bibinfo {author} {\bibfnamefont {X.}~\bibnamefont {Du}}, \bibinfo {author}
  {\bibfnamefont {A.}~\bibnamefont {Benson}}, \bibinfo {author} {\bibfnamefont
  {S.}~\bibnamefont {Kim}}, \bibinfo {author} {\bibfnamefont {F.}~\bibnamefont
  {Jiang}}, \bibinfo {author} {\bibfnamefont {F.-Y.}\ \bibnamefont
  {Cyr-Racine}}, \ and\ \bibinfo {author} {\bibfnamefont {M.}~\bibnamefont
  {Vogelsberger}},\ }\href {\doibase 10.1093/mnras/stac1094} {\bibfield
  {journal} {\bibinfo  {journal} {Mon. Not. Roy. Astron. Soc.}\ }\textbf
  {\bibinfo {volume} {513}},\ \bibinfo {pages} {4845} (\bibinfo {year}
  {2022})},\ \Eprint {http://arxiv.org/abs/2110.00259} {arXiv:2110.00259
  [astro-ph.CO]} \BibitemShut {NoStop}%
\bibitem [{\citenamefont {Turner}\ \emph {et~al.}(2021)\citenamefont {Turner},
  \citenamefont {Lovell}, \citenamefont {Zavala},\ and\ \citenamefont
  {Vogelsberger}}]{Turner:2020vlf}%
  \BibitemOpen
  \bibfield  {author} {\bibinfo {author} {\bibfnamefont {H.~C.}\ \bibnamefont
  {Turner}}, \bibinfo {author} {\bibfnamefont {M.~R.}\ \bibnamefont {Lovell}},
  \bibinfo {author} {\bibfnamefont {J.}~\bibnamefont {Zavala}}, \ and\ \bibinfo
  {author} {\bibfnamefont {M.}~\bibnamefont {Vogelsberger}},\ }\href {\doibase
  10.1093/mnras/stab1725} {\bibfield  {journal} {\bibinfo  {journal} {Mon. Not.
  Roy. Astron. Soc.}\ }\textbf {\bibinfo {volume} {505}},\ \bibinfo {pages}
  {5327} (\bibinfo {year} {2021})},\ \Eprint {http://arxiv.org/abs/2010.02924}
  {arXiv:2010.02924 [astro-ph.GA]} \BibitemShut {NoStop}%
\bibitem [{\citenamefont {Kahlhoefer}\ \emph {et~al.}(2019)\citenamefont
  {Kahlhoefer}, \citenamefont {Kaplinghat}, \citenamefont {Slatyer},\ and\
  \citenamefont {Wu}}]{Kahlhoefer:2019oyt}%
  \BibitemOpen
  \bibfield  {author} {\bibinfo {author} {\bibfnamefont {F.}~\bibnamefont
  {Kahlhoefer}}, \bibinfo {author} {\bibfnamefont {M.}~\bibnamefont
  {Kaplinghat}}, \bibinfo {author} {\bibfnamefont {T.~R.}\ \bibnamefont
  {Slatyer}}, \ and\ \bibinfo {author} {\bibfnamefont {C.-L.}\ \bibnamefont
  {Wu}},\ }\href {\doibase 10.1088/1475-7516/2019/12/010} {\bibfield  {journal}
  {\bibinfo  {journal} {JCAP}\ }\textbf {\bibinfo {volume} {12}},\ \bibinfo
  {pages} {010} (\bibinfo {year} {2019})},\ \Eprint
  {http://arxiv.org/abs/1904.10539} {arXiv:1904.10539 [astro-ph.GA]}
  \BibitemShut {NoStop}%
\bibitem [{\citenamefont {Jiang}\ \emph {et~al.}(2023)\citenamefont {Jiang}
  \emph {et~al.}}]{Jiang:2022aqw}%
  \BibitemOpen
  \bibfield  {author} {\bibinfo {author} {\bibfnamefont {F.}~\bibnamefont
  {Jiang}} \emph {et~al.},\ }\href {\doibase 10.1093/mnras/stad705} {\bibfield
  {journal} {\bibinfo  {journal} {Mon. Not. Roy. Astron. Soc.}\ }\textbf
  {\bibinfo {volume} {521}},\ \bibinfo {pages} {4630} (\bibinfo {year}
  {2023})},\ \Eprint {http://arxiv.org/abs/2206.12425} {arXiv:2206.12425
  [astro-ph.CO]} \BibitemShut {NoStop}%
\bibitem [{\citenamefont {Yang}\ \emph
  {et~al.}(2023{\natexlab{a}})\citenamefont {Yang}, \citenamefont {Nadler},\
  and\ \citenamefont {Yu}}]{Yang:2022mxl}%
  \BibitemOpen
  \bibfield  {author} {\bibinfo {author} {\bibfnamefont {D.}~\bibnamefont
  {Yang}}, \bibinfo {author} {\bibfnamefont {E.~O.}\ \bibnamefont {Nadler}}, \
  and\ \bibinfo {author} {\bibfnamefont {H.-B.}\ \bibnamefont {Yu}},\ }\href
  {\doibase 10.3847/1538-4357/acc73e} {\bibfield  {journal} {\bibinfo
  {journal} {Astrophys. J.}\ }\textbf {\bibinfo {volume} {949}},\ \bibinfo
  {pages} {67} (\bibinfo {year} {2023}{\natexlab{a}})},\ \Eprint
  {http://arxiv.org/abs/2211.13768} {arXiv:2211.13768 [astro-ph.GA]}
  \BibitemShut {NoStop}%
\bibitem [{\citenamefont {Pollack}\ \emph {et~al.}(2015)\citenamefont
  {Pollack}, \citenamefont {Spergel},\ and\ \citenamefont
  {Steinhardt}}]{Pollack:2014rja}%
  \BibitemOpen
  \bibfield  {author} {\bibinfo {author} {\bibfnamefont {J.}~\bibnamefont
  {Pollack}}, \bibinfo {author} {\bibfnamefont {D.~N.}\ \bibnamefont
  {Spergel}}, \ and\ \bibinfo {author} {\bibfnamefont {P.~J.}\ \bibnamefont
  {Steinhardt}},\ }\href {\doibase 10.1088/0004-637X/804/2/131} {\bibfield
  {journal} {\bibinfo  {journal} {Astrophys. J.}\ }\textbf {\bibinfo {volume}
  {804}},\ \bibinfo {pages} {131} (\bibinfo {year} {2015})},\ \Eprint
  {http://arxiv.org/abs/1501.00017} {arXiv:1501.00017 [astro-ph.CO]}
  \BibitemShut {NoStop}%
\bibitem [{\citenamefont {Outmezguine}\ \emph {et~al.}(2023)\citenamefont
  {Outmezguine}, \citenamefont {Boddy}, \citenamefont {Gad-Nasr}, \citenamefont
  {Kaplinghat},\ and\ \citenamefont {Sagunski}}]{Outmezguine:2022bhq}%
  \BibitemOpen
  \bibfield  {author} {\bibinfo {author} {\bibfnamefont {N.~J.}\ \bibnamefont
  {Outmezguine}}, \bibinfo {author} {\bibfnamefont {K.~K.}\ \bibnamefont
  {Boddy}}, \bibinfo {author} {\bibfnamefont {S.}~\bibnamefont {Gad-Nasr}},
  \bibinfo {author} {\bibfnamefont {M.}~\bibnamefont {Kaplinghat}}, \ and\
  \bibinfo {author} {\bibfnamefont {L.}~\bibnamefont {Sagunski}},\ }\href
  {\doibase 10.1093/mnras/stad1705} {\bibfield  {journal} {\bibinfo  {journal}
  {Mon. Not. Roy. Astron. Soc.}\ }\textbf {\bibinfo {volume} {523}},\ \bibinfo
  {pages} {4786} (\bibinfo {year} {2023})},\ \Eprint
  {http://arxiv.org/abs/2204.06568} {arXiv:2204.06568 [astro-ph.GA]}
  \BibitemShut {NoStop}%
\bibitem [{\citenamefont {Yang}\ \emph
  {et~al.}(2024{\natexlab{b}})\citenamefont {Yang}, \citenamefont {Jiang},
  \citenamefont {Benson}, \citenamefont {Zhong}, \citenamefont {Mace},
  \citenamefont {Du}, \citenamefont {Zeng}, \citenamefont {Peter},\ and\
  \citenamefont {Fischer}}]{Yang:2023stn}%
  \BibitemOpen
  \bibfield  {author} {\bibinfo {author} {\bibfnamefont {S.}~\bibnamefont
  {Yang}}, \bibinfo {author} {\bibfnamefont {F.}~\bibnamefont {Jiang}},
  \bibinfo {author} {\bibfnamefont {A.}~\bibnamefont {Benson}}, \bibinfo
  {author} {\bibfnamefont {Y.-M.}\ \bibnamefont {Zhong}}, \bibinfo {author}
  {\bibfnamefont {C.}~\bibnamefont {Mace}}, \bibinfo {author} {\bibfnamefont
  {X.}~\bibnamefont {Du}}, \bibinfo {author} {\bibfnamefont {Z.~C.}\
  \bibnamefont {Zeng}}, \bibinfo {author} {\bibfnamefont {A.~H.~G.}\
  \bibnamefont {Peter}}, \ and\ \bibinfo {author} {\bibfnamefont {M.~S.}\
  \bibnamefont {Fischer}},\ }\href {\doibase 10.1093/mnras/stae2038} {\bibfield
   {journal} {\bibinfo  {journal} {Mon. Not. Roy. Astron. Soc.}\ }\textbf
  {\bibinfo {volume} {533}},\ \bibinfo {pages} {4007} (\bibinfo {year}
  {2024}{\natexlab{b}})},\ \Eprint {http://arxiv.org/abs/2305.05067}
  {arXiv:2305.05067 [astro-ph.CO]} \BibitemShut {NoStop}%
\bibitem [{\citenamefont {Schneider}\ \emph {et~al.}(1992)\citenamefont
  {Schneider}, \citenamefont {Ehlers},\ and\ \citenamefont
  {Falco}}]{Schneider:1992bmb}%
  \BibitemOpen
  \bibfield  {author} {\bibinfo {author} {\bibfnamefont {P.}~\bibnamefont
  {Schneider}}, \bibinfo {author} {\bibfnamefont {J.}~\bibnamefont {Ehlers}}, \
  and\ \bibinfo {author} {\bibfnamefont {E.~E.}\ \bibnamefont {Falco}},\ }\href
  {\doibase 10.1007/978-3-662-03758-4} {\emph {\bibinfo {title} {{Gravitational
  Lenses}}}},\ Astronomy and Astrophysics Library\ (\bibinfo  {publisher}
  {Springer},\ \bibinfo {year} {1992})\BibitemShut {NoStop}%
\bibitem [{\citenamefont {Shin}\ \emph {et~al.}(2023)\citenamefont {Shin} \emph
  {et~al.}}]{Shin:2022crt}%
  \BibitemOpen
  \bibfield  {author} {\bibinfo {author} {\bibfnamefont {K.}~\bibnamefont
  {Shin}} \emph {et~al.},\ }\href {\doibase 10.3847/1538-4357/acaf06}
  {\bibfield  {journal} {\bibinfo  {journal} {Astrophys. J.}\ }\textbf
  {\bibinfo {volume} {944}},\ \bibinfo {pages} {105} (\bibinfo {year}
  {2023})},\ \Eprint {http://arxiv.org/abs/2207.14316} {arXiv:2207.14316
  [astro-ph.HE]} \BibitemShut {NoStop}%
\bibitem [{\citenamefont {Gupta}\ \emph {et~al.}(2025)\citenamefont {Gupta},
  \citenamefont {Beniamini}, \citenamefont {Kumar},\ and\ \citenamefont
  {Finkelstein}}]{Gupta:2025jyw}%
  \BibitemOpen
  \bibfield  {author} {\bibinfo {author} {\bibfnamefont {O.}~\bibnamefont
  {Gupta}}, \bibinfo {author} {\bibfnamefont {P.}~\bibnamefont {Beniamini}},
  \bibinfo {author} {\bibfnamefont {P.}~\bibnamefont {Kumar}}, \ and\ \bibinfo
  {author} {\bibfnamefont {S.~L.}\ \bibnamefont {Finkelstein}},\ }\href
  {\doibase 10.3847/1538-4357/add14c} {\bibfield  {journal} {\bibinfo
  {journal} {Astrophys. J.}\ }\textbf {\bibinfo {volume} {986}},\ \bibinfo
  {pages} {100} (\bibinfo {year} {2025})},\ \Eprint
  {http://arxiv.org/abs/2501.09810} {arXiv:2501.09810 [astro-ph.HE]}
  \BibitemShut {NoStop}%
\bibitem [{\citenamefont {Lei}\ \emph {et~al.}(2025)\citenamefont {Lei},
  \citenamefont {Wang},\ and\ \citenamefont {Deng}}]{Lei:2025wyw}%
  \BibitemOpen
  \bibfield  {author} {\bibinfo {author} {\bibfnamefont {Q.-Z.}\ \bibnamefont
  {Lei}}, \bibinfo {author} {\bibfnamefont {X.-Z.}\ \bibnamefont {Wang}}, \
  and\ \bibinfo {author} {\bibfnamefont {C.-M.}\ \bibnamefont {Deng}},\ }\href
  {\doibase 10.3847/1538-4357/adf646} {\bibfield  {journal} {\bibinfo
  {journal} {Astrophys. J.}\ }\textbf {\bibinfo {volume} {990}},\ \bibinfo
  {pages} {175} (\bibinfo {year} {2025})},\ \Eprint
  {http://arxiv.org/abs/2507.23122} {arXiv:2507.23122 [astro-ph.HE]}
  \BibitemShut {NoStop}%
\bibitem [{\citenamefont {Chen}\ \emph {et~al.}(2024)\citenamefont {Chen},
  \citenamefont {Jia}, \citenamefont {Dong},\ and\ \citenamefont
  {Wang}}]{Chen:2024fne}%
  \BibitemOpen
  \bibfield  {author} {\bibinfo {author} {\bibfnamefont {J.~H.}\ \bibnamefont
  {Chen}}, \bibinfo {author} {\bibfnamefont {X.~D.}\ \bibnamefont {Jia}},
  \bibinfo {author} {\bibfnamefont {X.~F.}\ \bibnamefont {Dong}}, \ and\
  \bibinfo {author} {\bibfnamefont {F.~Y.}\ \bibnamefont {Wang}},\ }\href
  {\doibase 10.3847/2041-8213/ad7b39} {\bibfield  {journal} {\bibinfo
  {journal} {Astrophys. J. Lett.}\ }\textbf {\bibinfo {volume} {973}},\
  \bibinfo {pages} {L54} (\bibinfo {year} {2024})},\ \Eprint
  {http://arxiv.org/abs/2406.03672} {arXiv:2406.03672 [astro-ph.HE]}
  \BibitemShut {NoStop}%
\bibitem [{\citenamefont {Luo}\ \emph {et~al.}(2020)\citenamefont {Luo},
  \citenamefont {Men}, \citenamefont {Lee}, \citenamefont {Wang}, \citenamefont
  {Lorimer},\ and\ \citenamefont {Zhang}}]{Luo:2020wfx}%
  \BibitemOpen
  \bibfield  {author} {\bibinfo {author} {\bibfnamefont {R.}~\bibnamefont
  {Luo}}, \bibinfo {author} {\bibfnamefont {Y.}~\bibnamefont {Men}}, \bibinfo
  {author} {\bibfnamefont {K.}~\bibnamefont {Lee}}, \bibinfo {author}
  {\bibfnamefont {W.}~\bibnamefont {Wang}}, \bibinfo {author} {\bibfnamefont
  {D.~R.}\ \bibnamefont {Lorimer}}, \ and\ \bibinfo {author} {\bibfnamefont
  {B.}~\bibnamefont {Zhang}},\ }\href {\doibase 10.1093/mnras/staa704}
  {\bibfield  {journal} {\bibinfo  {journal} {Mon. Not. Roy. Astron. Soc.}\
  }\textbf {\bibinfo {volume} {494}},\ \bibinfo {pages} {665} (\bibinfo {year}
  {2020})},\ \Eprint {http://arxiv.org/abs/2003.04848} {arXiv:2003.04848
  [astro-ph.HE]} \BibitemShut {NoStop}%
\bibitem [{\citenamefont {Connor}\ and\ \citenamefont
  {Ravi}(2023)}]{Connor:2022bwl}%
  \BibitemOpen
  \bibfield  {author} {\bibinfo {author} {\bibfnamefont {L.}~\bibnamefont
  {Connor}}\ and\ \bibinfo {author} {\bibfnamefont {V.}~\bibnamefont {Ravi}},\
  }\href {\doibase 10.1093/mnras/stad667} {\bibfield  {journal} {\bibinfo
  {journal} {Mon. Not. Roy. Astron. Soc.}\ }\textbf {\bibinfo {volume} {521}},\
  \bibinfo {pages} {4024} (\bibinfo {year} {2023})},\ \Eprint
  {http://arxiv.org/abs/2206.14310} {arXiv:2206.14310 [astro-ph.CO]}
  \BibitemShut {NoStop}%
\bibitem [{\citenamefont {Lin}\ \emph {et~al.}(2022)\citenamefont {Lin} \emph
  {et~al.}}]{Lin:2022wgp}%
  \BibitemOpen
  \bibfield  {author} {\bibinfo {author} {\bibfnamefont {H.-H.}\ \bibnamefont
  {Lin}} \emph {et~al.},\ }\href {\doibase 10.1088/1538-3873/ac8f71} {\bibfield
   {journal} {\bibinfo  {journal} {Publ. Astron. Soc. Pac.}\ }\textbf {\bibinfo
  {volume} {134}},\ \bibinfo {pages} {094106} (\bibinfo {year} {2022})},\
  \Eprint {http://arxiv.org/abs/2206.08983} {arXiv:2206.08983 [astro-ph.IM]}
  \BibitemShut {NoStop}%
\bibitem [{\citenamefont {Braun}\ \emph {et~al.}(2019)\citenamefont {Braun},
  \citenamefont {Bonaldi}, \citenamefont {Bourke}, \citenamefont {Keane},\ and\
  \citenamefont {Wagg}}]{braun2019}%
  \BibitemOpen
  \bibfield  {author} {\bibinfo {author} {\bibfnamefont {R.}~\bibnamefont
  {Braun}}, \bibinfo {author} {\bibfnamefont {A.}~\bibnamefont {Bonaldi}},
  \bibinfo {author} {\bibfnamefont {T.}~\bibnamefont {Bourke}}, \bibinfo
  {author} {\bibfnamefont {E.}~\bibnamefont {Keane}}, \ and\ \bibinfo {author}
  {\bibfnamefont {J.}~\bibnamefont {Wagg}},\ }\href
  {https://arxiv.org/abs/1912.12699} {\enquote {\bibinfo {title} {Anticipated
  performance of the square kilometre array -- phase 1 (ska1)},}\ } (\bibinfo
  {year} {2019}),\ \Eprint {http://arxiv.org/abs/1912.12699} {arXiv:1912.12699
  [astro-ph.IM]} \BibitemShut {NoStop}%
\bibitem [{\citenamefont {Collaboration}\ \emph {et~al.}(2018)\citenamefont
  {Collaboration}, \citenamefont {Amiri}, \citenamefont {Bandura},
  \citenamefont {Berger}, \citenamefont {Bhardwaj}, \citenamefont {Boyce},
  \citenamefont {Boyle}, \citenamefont {Brar}, \citenamefont {Burhanpurkar},
  \citenamefont {Chawla}, \citenamefont {Chowdhury}, \citenamefont {Cliche},
  \citenamefont {Cranmer}, \citenamefont {Cubranic}, \citenamefont {Deng},
  \citenamefont {Denman}, \citenamefont {Dobbs}, \citenamefont {Fandino},
  \citenamefont {Fonseca}, \citenamefont {Gaensler}, \citenamefont {Giri},
  \citenamefont {Gilbert}, \citenamefont {Good}, \citenamefont {Guliani},
  \citenamefont {Halpern}, \citenamefont {Hinshaw}, \citenamefont {Höfer},
  \citenamefont {Josephy}, \citenamefont {Kaspi}, \citenamefont {Landecker},
  \citenamefont {Lang}, \citenamefont {Liao}, \citenamefont {Masui},
  \citenamefont {Mena-Parra}, \citenamefont {Naidu}, \citenamefont {Newburgh},
  \citenamefont {Ng}, \citenamefont {Patel}, \citenamefont {Pen}, \citenamefont
  {Pinsonneault-Marotte}, \citenamefont {Pleunis}, \citenamefont {Ravandi},
  \citenamefont {Ransom}, \citenamefont {Renard}, \citenamefont {Scholz},
  \citenamefont {Sigurdson}, \citenamefont {Siegel}, \citenamefont {Smith},
  \citenamefont {Stairs}, \citenamefont {Tendulkar}, \citenamefont
  {Vanderlinde},\ and\ \citenamefont {Wiebe}}]{Amiri_2018}%
  \BibitemOpen
  \bibfield  {author} {\bibinfo {author} {\bibfnamefont {T.~C.}\ \bibnamefont
  {Collaboration}}, \bibinfo {author} {\bibfnamefont {M.}~\bibnamefont
  {Amiri}}, \bibinfo {author} {\bibfnamefont {K.}~\bibnamefont {Bandura}},
  \bibinfo {author} {\bibfnamefont {P.}~\bibnamefont {Berger}}, \bibinfo
  {author} {\bibfnamefont {M.}~\bibnamefont {Bhardwaj}}, \bibinfo {author}
  {\bibfnamefont {M.~M.}\ \bibnamefont {Boyce}}, \bibinfo {author}
  {\bibfnamefont {P.~J.}\ \bibnamefont {Boyle}}, \bibinfo {author}
  {\bibfnamefont {C.}~\bibnamefont {Brar}}, \bibinfo {author} {\bibfnamefont
  {M.}~\bibnamefont {Burhanpurkar}}, \bibinfo {author} {\bibfnamefont
  {P.}~\bibnamefont {Chawla}}, \bibinfo {author} {\bibfnamefont
  {J.}~\bibnamefont {Chowdhury}}, \bibinfo {author} {\bibfnamefont {J.-F.}\
  \bibnamefont {Cliche}}, \bibinfo {author} {\bibfnamefont {M.~D.}\
  \bibnamefont {Cranmer}}, \bibinfo {author} {\bibfnamefont {D.}~\bibnamefont
  {Cubranic}}, \bibinfo {author} {\bibfnamefont {M.}~\bibnamefont {Deng}},
  \bibinfo {author} {\bibfnamefont {N.}~\bibnamefont {Denman}}, \bibinfo
  {author} {\bibfnamefont {M.}~\bibnamefont {Dobbs}}, \bibinfo {author}
  {\bibfnamefont {M.}~\bibnamefont {Fandino}}, \bibinfo {author} {\bibfnamefont
  {E.}~\bibnamefont {Fonseca}}, \bibinfo {author} {\bibfnamefont {B.~M.}\
  \bibnamefont {Gaensler}}, \bibinfo {author} {\bibfnamefont {U.}~\bibnamefont
  {Giri}}, \bibinfo {author} {\bibfnamefont {A.~J.}\ \bibnamefont {Gilbert}},
  \bibinfo {author} {\bibfnamefont {D.~C.}\ \bibnamefont {Good}}, \bibinfo
  {author} {\bibfnamefont {S.}~\bibnamefont {Guliani}}, \bibinfo {author}
  {\bibfnamefont {M.}~\bibnamefont {Halpern}}, \bibinfo {author} {\bibfnamefont
  {G.}~\bibnamefont {Hinshaw}}, \bibinfo {author} {\bibfnamefont
  {C.}~\bibnamefont {Höfer}}, \bibinfo {author} {\bibfnamefont
  {A.}~\bibnamefont {Josephy}}, \bibinfo {author} {\bibfnamefont {V.~M.}\
  \bibnamefont {Kaspi}}, \bibinfo {author} {\bibfnamefont {T.~L.}\ \bibnamefont
  {Landecker}}, \bibinfo {author} {\bibfnamefont {D.}~\bibnamefont {Lang}},
  \bibinfo {author} {\bibfnamefont {H.}~\bibnamefont {Liao}}, \bibinfo {author}
  {\bibfnamefont {K.~W.}\ \bibnamefont {Masui}}, \bibinfo {author}
  {\bibfnamefont {J.}~\bibnamefont {Mena-Parra}}, \bibinfo {author}
  {\bibfnamefont {A.}~\bibnamefont {Naidu}}, \bibinfo {author} {\bibfnamefont
  {L.~B.}\ \bibnamefont {Newburgh}}, \bibinfo {author} {\bibfnamefont
  {C.}~\bibnamefont {Ng}}, \bibinfo {author} {\bibfnamefont {C.}~\bibnamefont
  {Patel}}, \bibinfo {author} {\bibfnamefont {U.-L.}\ \bibnamefont {Pen}},
  \bibinfo {author} {\bibfnamefont {T.}~\bibnamefont {Pinsonneault-Marotte}},
  \bibinfo {author} {\bibfnamefont {Z.}~\bibnamefont {Pleunis}}, \bibinfo
  {author} {\bibfnamefont {M.~R.}\ \bibnamefont {Ravandi}}, \bibinfo {author}
  {\bibfnamefont {S.~M.}\ \bibnamefont {Ransom}}, \bibinfo {author}
  {\bibfnamefont {A.}~\bibnamefont {Renard}}, \bibinfo {author} {\bibfnamefont
  {P.}~\bibnamefont {Scholz}}, \bibinfo {author} {\bibfnamefont
  {K.}~\bibnamefont {Sigurdson}}, \bibinfo {author} {\bibfnamefont {S.~R.}\
  \bibnamefont {Siegel}}, \bibinfo {author} {\bibfnamefont {K.~M.}\
  \bibnamefont {Smith}}, \bibinfo {author} {\bibfnamefont {I.~H.}\ \bibnamefont
  {Stairs}}, \bibinfo {author} {\bibfnamefont {S.~P.}\ \bibnamefont
  {Tendulkar}}, \bibinfo {author} {\bibfnamefont {K.}~\bibnamefont
  {Vanderlinde}}, \ and\ \bibinfo {author} {\bibfnamefont {D.~V.}\ \bibnamefont
  {Wiebe}},\ }\href {\doibase 10.3847/1538-4357/aad188} {\bibfield  {journal}
  {\bibinfo  {journal} {The Astrophysical Journal}\ }\textbf {\bibinfo {volume}
  {863}},\ \bibinfo {pages} {48} (\bibinfo {year} {2018})}\BibitemShut
  {NoStop}%
\bibitem [{\citenamefont {Normile}(2025)}]{Normile2025BURSTT}%
  \BibitemOpen
  \bibfield  {author} {\bibinfo {author} {\bibfnamefont {D.}~\bibnamefont
  {Normile}},\ }\href
  {https://www.science.org/content/article/new-kind-telescope-set-search-mysterious-fast-radio-bursts}
  {\bibfield  {journal} {\bibinfo  {journal} {Science}\ } (\bibinfo {year}
  {2025})},\ \bibinfo {note} {news article}\BibitemShut {NoStop}%
\bibitem [{\citenamefont {{Jiang}}\ \emph {et~al.}(2024)\citenamefont
  {{Jiang}}, \citenamefont {{Chen}}, \citenamefont {{Gan}}, \citenamefont
  {{Sun}}, \citenamefont {{Zhu}}, \citenamefont {{Li}}, \citenamefont {{Zhu}},
  \citenamefont {{Wu}}, \citenamefont {{Chen}}, \citenamefont {{Zhang}},\ and\
  \citenamefont {{An}}}]{2024Jiang}%
  \BibitemOpen
  \bibfield  {author} {\bibinfo {author} {\bibfnamefont {P.}~\bibnamefont
  {{Jiang}}}, \bibinfo {author} {\bibfnamefont {R.}~\bibnamefont {{Chen}}},
  \bibinfo {author} {\bibfnamefont {H.}~\bibnamefont {{Gan}}}, \bibinfo
  {author} {\bibfnamefont {J.}~\bibnamefont {{Sun}}}, \bibinfo {author}
  {\bibfnamefont {B.}~\bibnamefont {{Zhu}}}, \bibinfo {author} {\bibfnamefont
  {H.}~\bibnamefont {{Li}}}, \bibinfo {author} {\bibfnamefont {W.}~\bibnamefont
  {{Zhu}}}, \bibinfo {author} {\bibfnamefont {J.}~\bibnamefont {{Wu}}},
  \bibinfo {author} {\bibfnamefont {X.}~\bibnamefont {{Chen}}}, \bibinfo
  {author} {\bibfnamefont {H.}~\bibnamefont {{Zhang}}}, \ and\ \bibinfo
  {author} {\bibfnamefont {T.}~\bibnamefont {{An}}},\ }\href {\doibase
  10.61977/ati2024012} {\bibfield  {journal} {\bibinfo  {journal} {Astronomical
  Techniques and Instruments}\ }\textbf {\bibinfo {volume} {1}},\ \bibinfo
  {pages} {84} (\bibinfo {year} {2024})},\ \Eprint
  {http://arxiv.org/abs/2408.12826} {arXiv:2408.12826 [astro-ph.IM]}
  \BibitemShut {NoStop}%
\bibitem [{\citenamefont {Kader}\ \emph {et~al.}(2022)\citenamefont {Kader}
  \emph {et~al.}}]{CHIMEFRB:2022xzl}%
  \BibitemOpen
  \bibfield  {author} {\bibinfo {author} {\bibfnamefont {Z.}~\bibnamefont
  {Kader}} \emph {et~al.} (\bibinfo {collaboration} {CHIME/FRB}),\ }\href
  {\doibase 10.1103/PhysRevD.106.043016} {\bibfield  {journal} {\bibinfo
  {journal} {Phys. Rev. D}\ }\textbf {\bibinfo {volume} {106}},\ \bibinfo
  {pages} {043016} (\bibinfo {year} {2022})},\ \Eprint
  {http://arxiv.org/abs/2204.06014} {arXiv:2204.06014 [astro-ph.HE]}
  \BibitemShut {NoStop}%
\bibitem [{\citenamefont {Yang}\ \emph
  {et~al.}(2023{\natexlab{b}})\citenamefont {Yang}, \citenamefont {Du},
  \citenamefont {Zeng}, \citenamefont {Benson}, \citenamefont {Jiang},
  \citenamefont {Nadler},\ and\ \citenamefont {Peter}}]{Yang:2022zkd}%
  \BibitemOpen
  \bibfield  {author} {\bibinfo {author} {\bibfnamefont {S.}~\bibnamefont
  {Yang}}, \bibinfo {author} {\bibfnamefont {X.}~\bibnamefont {Du}}, \bibinfo
  {author} {\bibfnamefont {Z.~C.}\ \bibnamefont {Zeng}}, \bibinfo {author}
  {\bibfnamefont {A.}~\bibnamefont {Benson}}, \bibinfo {author} {\bibfnamefont
  {F.}~\bibnamefont {Jiang}}, \bibinfo {author} {\bibfnamefont {E.~O.}\
  \bibnamefont {Nadler}}, \ and\ \bibinfo {author} {\bibfnamefont {A.~H.~G.}\
  \bibnamefont {Peter}},\ }\href {\doibase 10.3847/1538-4357/acbd49} {\bibfield
   {journal} {\bibinfo  {journal} {Astrophys. J.}\ }\textbf {\bibinfo {volume}
  {946}},\ \bibinfo {pages} {47} (\bibinfo {year} {2023}{\natexlab{b}})},\
  \Eprint {http://arxiv.org/abs/2205.02957} {arXiv:2205.02957 [astro-ph.CO]}
  \BibitemShut {NoStop}%
\bibitem [{\citenamefont {Yang}\ and\ \citenamefont {Yu}(2022)}]{Yang:2022hkm}%
  \BibitemOpen
  \bibfield  {author} {\bibinfo {author} {\bibfnamefont {D.}~\bibnamefont
  {Yang}}\ and\ \bibinfo {author} {\bibfnamefont {H.-B.}\ \bibnamefont {Yu}},\
  }\href {\doibase 10.1088/1475-7516/2022/09/077} {\bibfield  {journal}
  {\bibinfo  {journal} {JCAP}\ }\textbf {\bibinfo {volume} {09}},\ \bibinfo
  {pages} {077} (\bibinfo {year} {2022})},\ \Eprint
  {http://arxiv.org/abs/2205.03392} {arXiv:2205.03392 [astro-ph.CO]}
  \BibitemShut {NoStop}%
\bibitem [{\citenamefont {Zhong}\ \emph {et~al.}(2023)\citenamefont {Zhong},
  \citenamefont {Yang},\ and\ \citenamefont {Yu}}]{Zhong:2023yzk}%
  \BibitemOpen
  \bibfield  {author} {\bibinfo {author} {\bibfnamefont {Y.-M.}\ \bibnamefont
  {Zhong}}, \bibinfo {author} {\bibfnamefont {D.}~\bibnamefont {Yang}}, \ and\
  \bibinfo {author} {\bibfnamefont {H.-B.}\ \bibnamefont {Yu}},\ }\href
  {\doibase 10.1093/mnras/stad2765} {\bibfield  {journal} {\bibinfo  {journal}
  {Mon. Not. Roy. Astron. Soc.}\ }\textbf {\bibinfo {volume} {526}},\ \bibinfo
  {pages} {758} (\bibinfo {year} {2023})},\ \Eprint
  {http://arxiv.org/abs/2306.08028} {arXiv:2306.08028 [astro-ph.CO]}
  \BibitemShut {NoStop}%
\bibitem [{\citenamefont {Foreman-Mackey}\ \emph {et~al.}(2013)\citenamefont
  {Foreman-Mackey}, \citenamefont {Hogg}, \citenamefont {Lang},\ and\
  \citenamefont {Goodman}}]{Foreman-Mackey:2012any}%
  \BibitemOpen
  \bibfield  {author} {\bibinfo {author} {\bibfnamefont {D.}~\bibnamefont
  {Foreman-Mackey}}, \bibinfo {author} {\bibfnamefont {D.~W.}\ \bibnamefont
  {Hogg}}, \bibinfo {author} {\bibfnamefont {D.}~\bibnamefont {Lang}}, \ and\
  \bibinfo {author} {\bibfnamefont {J.}~\bibnamefont {Goodman}},\ }\href
  {\doibase 10.1086/670067} {\bibfield  {journal} {\bibinfo  {journal} {Publ.
  Astron. Soc. Pac.}\ }\textbf {\bibinfo {volume} {125}},\ \bibinfo {pages}
  {306} (\bibinfo {year} {2013})},\ \Eprint {http://arxiv.org/abs/1202.3665}
  {arXiv:1202.3665 [astro-ph.IM]} \BibitemShut {NoStop}%
\bibitem [{\citenamefont {Spitzer}(2014)}]{spitzer2014dynamical}%
  \BibitemOpen
  \bibfield  {author} {\bibinfo {author} {\bibfnamefont {L.}~\bibnamefont
  {Spitzer}},\ }\href {https://books.google.com.hk/books?id=fPj_AwAAQBAJ}
  {\emph {\bibinfo {title} {Dynamical Evolution of Globular Clusters}}},\
  Princeton Series in Astrophysics\ (\bibinfo  {publisher} {Princeton
  University Press},\ \bibinfo {year} {2014})\BibitemShut {NoStop}%
\bibitem [{\citenamefont {Grossman}\ and\ \citenamefont
  {Nowak}(1994)}]{Grossman:1994je}%
  \BibitemOpen
  \bibfield  {author} {\bibinfo {author} {\bibfnamefont {S.~A.}\ \bibnamefont
  {Grossman}}\ and\ \bibinfo {author} {\bibfnamefont {M.~A.}\ \bibnamefont
  {Nowak}},\ }\href {\doibase 10.1086/174836} {\bibfield  {journal} {\bibinfo
  {journal} {Astrophys. J.}\ }\textbf {\bibinfo {volume} {435}},\ \bibinfo
  {pages} {548} (\bibinfo {year} {1994})},\ \Eprint
  {http://arxiv.org/abs/astro-ph/9401047} {arXiv:astro-ph/9401047} \BibitemShut
  {NoStop}%
\bibitem [{\citenamefont {Ji}\ \emph {et~al.}(2018)\citenamefont {Ji},
  \citenamefont {Kovetz},\ and\ \citenamefont {Kamionkowski}}]{Ji:2018rvg}%
  \BibitemOpen
  \bibfield  {author} {\bibinfo {author} {\bibfnamefont {L.}~\bibnamefont
  {Ji}}, \bibinfo {author} {\bibfnamefont {E.~D.}\ \bibnamefont {Kovetz}}, \
  and\ \bibinfo {author} {\bibfnamefont {M.}~\bibnamefont {Kamionkowski}},\
  }\href {\doibase 10.1103/PhysRevD.98.123523} {\bibfield  {journal} {\bibinfo
  {journal} {Phys. Rev. D}\ }\textbf {\bibinfo {volume} {98}},\ \bibinfo
  {pages} {123523} (\bibinfo {year} {2018})},\ \Eprint
  {http://arxiv.org/abs/1809.09627} {arXiv:1809.09627 [astro-ph.CO]}
  \BibitemShut {NoStop}%
\bibitem [{\citenamefont {Goobar}\ \emph {et~al.}(2025)\citenamefont {Goobar},
  \citenamefont {Johansson},\ and\ \citenamefont {Carracedo}}]{Goobar:2024dzh}%
  \BibitemOpen
  \bibfield  {author} {\bibinfo {author} {\bibfnamefont {A.}~\bibnamefont
  {Goobar}}, \bibinfo {author} {\bibfnamefont {J.}~\bibnamefont {Johansson}}, \
  and\ \bibinfo {author} {\bibfnamefont {A.~S.}\ \bibnamefont {Carracedo}},\
  }\href {\doibase 10.1098/rsta.2024.0123} {\bibfield  {journal} {\bibinfo
  {journal} {Phil. Trans. Roy. Soc. Lond. A}\ }\textbf {\bibinfo {volume}
  {383}},\ \bibinfo {pages} {20240123} (\bibinfo {year} {2025})},\ \Eprint
  {http://arxiv.org/abs/2406.13519} {arXiv:2406.13519 [astro-ph.CO]}
  \BibitemShut {NoStop}%
\bibitem [{\citenamefont {Levan}\ \emph {et~al.}(2025)\citenamefont {Levan},
  \citenamefont {Gompertz}, \citenamefont {Smith}, \citenamefont {Ravasio},
  \citenamefont {Lamb},\ and\ \citenamefont {Tanvir}}]{Levan:2025ool}%
  \BibitemOpen
  \bibfield  {author} {\bibinfo {author} {\bibfnamefont {A.~J.}\ \bibnamefont
  {Levan}}, \bibinfo {author} {\bibfnamefont {B.~P.}\ \bibnamefont {Gompertz}},
  \bibinfo {author} {\bibfnamefont {G.~P.}\ \bibnamefont {Smith}}, \bibinfo
  {author} {\bibfnamefont {M.~E.}\ \bibnamefont {Ravasio}}, \bibinfo {author}
  {\bibfnamefont {G.}~\bibnamefont {Lamb}}, \ and\ \bibinfo {author}
  {\bibfnamefont {N.~R.}\ \bibnamefont {Tanvir}},\ }\href {\doibase
  10.1098/rsta.2024.0122} {\bibfield  {journal} {\bibinfo  {journal} {Phil.
  Trans. Roy. Soc. Lond. A}\ }\textbf {\bibinfo {volume} {383}},\ \bibinfo
  {pages} {20240122} (\bibinfo {year} {2025})},\ \Eprint
  {http://arxiv.org/abs/2503.19977} {arXiv:2503.19977 [astro-ph.HE]}
  \BibitemShut {NoStop}%
\bibitem [{\citenamefont {Suyu}\ \emph {et~al.}(2024)\citenamefont {Suyu},
  \citenamefont {Goobar}, \citenamefont {Collett}, \citenamefont {More},\ and\
  \citenamefont {Vernardos}}]{Suyu:2023jue}%
  \BibitemOpen
  \bibfield  {author} {\bibinfo {author} {\bibfnamefont {S.~H.}\ \bibnamefont
  {Suyu}}, \bibinfo {author} {\bibfnamefont {A.}~\bibnamefont {Goobar}},
  \bibinfo {author} {\bibfnamefont {T.}~\bibnamefont {Collett}}, \bibinfo
  {author} {\bibfnamefont {A.}~\bibnamefont {More}}, \ and\ \bibinfo {author}
  {\bibfnamefont {G.}~\bibnamefont {Vernardos}},\ }\href {\doibase
  10.1007/s11214-024-01044-7} {\bibfield  {journal} {\bibinfo  {journal} {Space
  Sci. Rev.}\ }\textbf {\bibinfo {volume} {220}},\ \bibinfo {pages} {13}
  (\bibinfo {year} {2024})},\ \Eprint {http://arxiv.org/abs/2301.07729}
  {arXiv:2301.07729 [astro-ph.CO]} \BibitemShut {NoStop}%
\bibitem [{\citenamefont {Yue}\ \emph {et~al.}(2022)\citenamefont {Yue},
  \citenamefont {Fan}, \citenamefont {Yang},\ and\ \citenamefont
  {Wang}}]{Yue:2021nwt}%
  \BibitemOpen
  \bibfield  {author} {\bibinfo {author} {\bibfnamefont {M.}~\bibnamefont
  {Yue}}, \bibinfo {author} {\bibfnamefont {X.}~\bibnamefont {Fan}}, \bibinfo
  {author} {\bibfnamefont {J.}~\bibnamefont {Yang}}, \ and\ \bibinfo {author}
  {\bibfnamefont {F.}~\bibnamefont {Wang}},\ }\href {\doibase
  10.3847/1538-4357/ac409b} {\bibfield  {journal} {\bibinfo  {journal}
  {Astrophys. J.}\ }\textbf {\bibinfo {volume} {925}},\ \bibinfo {pages} {169}
  (\bibinfo {year} {2022})},\ \Eprint {http://arxiv.org/abs/2112.02821}
  {arXiv:2112.02821 [astro-ph.GA]} \BibitemShut {NoStop}%
\bibitem [{\citenamefont {Geng}\ \emph {et~al.}(2021)\citenamefont {Geng},
  \citenamefont {Cao}, \citenamefont {Liu}, \citenamefont {Liu}, \citenamefont
  {Biesiada},\ and\ \citenamefont {Lian}}]{geng:2021}%
  \BibitemOpen
  \bibfield  {author} {\bibinfo {author} {\bibfnamefont {S.}~\bibnamefont
  {Geng}}, \bibinfo {author} {\bibfnamefont {S.}~\bibnamefont {Cao}}, \bibinfo
  {author} {\bibfnamefont {Y.}~\bibnamefont {Liu}}, \bibinfo {author}
  {\bibfnamefont {T.}~\bibnamefont {Liu}}, \bibinfo {author} {\bibfnamefont
  {M.}~\bibnamefont {Biesiada}}, \ and\ \bibinfo {author} {\bibfnamefont
  {Y.}~\bibnamefont {Lian}},\ }\href {\doibase 10.1093/mnras/stab519}
  {\bibfield  {journal} {\bibinfo  {journal} {Monthly Notices of the Royal
  Astronomical Society}\ }\textbf {\bibinfo {volume} {503}},\ \bibinfo {pages}
  {1319} (\bibinfo {year} {2021})},\ \Eprint
  {http://arxiv.org/abs/https://academic.oup.com/mnras/article-pdf/503/1/1319/36636581/stab519.pdf}
  {https://academic.oup.com/mnras/article-pdf/503/1/1319/36636581/stab519.pdf}
  \BibitemShut {NoStop}%
\bibitem [{\citenamefont {Gilman}\ \emph {et~al.}(2020)\citenamefont {Gilman},
  \citenamefont {Birrer}, \citenamefont {Nierenberg}, \citenamefont {Treu},
  \citenamefont {Du},\ and\ \citenamefont {Benson}}]{Gilman:2019nap}%
  \BibitemOpen
  \bibfield  {author} {\bibinfo {author} {\bibfnamefont {D.}~\bibnamefont
  {Gilman}}, \bibinfo {author} {\bibfnamefont {S.}~\bibnamefont {Birrer}},
  \bibinfo {author} {\bibfnamefont {A.}~\bibnamefont {Nierenberg}}, \bibinfo
  {author} {\bibfnamefont {T.}~\bibnamefont {Treu}}, \bibinfo {author}
  {\bibfnamefont {X.}~\bibnamefont {Du}}, \ and\ \bibinfo {author}
  {\bibfnamefont {A.}~\bibnamefont {Benson}},\ }\href {\doibase
  10.1093/mnras/stz3480} {\bibfield  {journal} {\bibinfo  {journal} {Mon. Not.
  Roy. Astron. Soc.}\ }\textbf {\bibinfo {volume} {491}},\ \bibinfo {pages}
  {6077} (\bibinfo {year} {2020})},\ \Eprint {http://arxiv.org/abs/1908.06983}
  {arXiv:1908.06983 [astro-ph.CO]} \BibitemShut {NoStop}%
\bibitem [{\citenamefont {Lepage}(2021)}]{Lepage:2020tgj}%
  \BibitemOpen
  \bibfield  {author} {\bibinfo {author} {\bibfnamefont {G.~P.}\ \bibnamefont
  {Lepage}},\ }\href {\doibase 10.1016/j.jcp.2021.110386} {\bibfield  {journal}
  {\bibinfo  {journal} {J. Comput. Phys.}\ }\textbf {\bibinfo {volume} {439}},\
  \bibinfo {pages} {110386} (\bibinfo {year} {2021})},\ \Eprint
  {http://arxiv.org/abs/2009.05112} {arXiv:2009.05112 [physics.comp-ph]}
  \BibitemShut {NoStop}%
\end{thebibliography}%
\bibliographystyle{apsrev4-2}

%\clearpage
\appendix

\onecolumngrid
\newpage
\begin{center}
{\bf \Large Supplemental Material}
\end{center}

\setcounter{equation}{0}
\setcounter{figure}{0}
\setcounter{table}{0}
\makeatletter
\renewcommand{\theequation}{S\arabic{equation}}
\renewcommand{\thefigure}{S\arabic{figure}}
\renewcommand{\thetable}{S\arabic{table}}

\section{Details of Lensing Properties}

Under the thin-lens approximation, the gravitational lensing properties are encoded in the Fermat potential
\begin{equation}
\Phi(\boldsymbol{\theta},\boldsymbol{\beta})= \frac{1}{2}|\boldsymbol{\theta} -\boldsymbol{\beta}|^2 - \psi(\boldsymbol{\theta}),
\end{equation}
where $\boldsymbol{\beta}$ and $\boldsymbol{\theta}$ are the angular position vectors on the source and lens planes, respectively, and
\begin{equation}
\psi(\boldsymbol{\theta}) =\frac{4 G D_{ls}D_l}{c^2 D_s } \int d^2 \boldsymbol{\theta}^\prime \Sigma(\boldsymbol{\theta}^\prime D_l) \ln|\boldsymbol{\theta}-\boldsymbol{\theta}^\prime|.
\end{equation}
Here $D_{ls}$, $D_l$, and $D_s$ are the angular diameter distances from the lens to the source, from the observer to the lens, and from the observer to the source, respectively, $\Sigma(\boldsymbol{\theta})$ is the projected surface mass density of the lens, and $c$ is the speed of light. Image positions correspond to stationary points of the Fermat potential, satisfying the lens equation
\begin{equation}
    \boldsymbol{\beta} = \boldsymbol{\theta}- \nabla_{\boldsymbol{\theta}}\psi(\boldsymbol{\theta}).
\label{eq:le}
\end{equation}
The time delay and magnification of the $i$-th image at $\boldsymbol{\theta}_i$ are given by
\begin{align}
    \Delta t_i ={}& (1+z_l) \frac{D_{ls} D_l}{c D_s} \Phi(\boldsymbol{\theta}_i,\boldsymbol{\beta}),\\
    \mu_i ={}& \frac{1}{\left|\det\left(\mathrm{Hess}\,\Phi(\boldsymbol{\theta}_i,\boldsymbol{\beta})\right)\right|},
\end{align}
where ``Hess'' denotes the Hessian matrix. The critical curves are obtained by solving $\mu_i^{-1}=0$, and the caustics follow by mapping these curves back to the source plane via~\eqref{le}.

For an axially symmetric lens, the lens equation simplifies to
\begin{equation}
    \boldsymbol{\beta} = \boldsymbol{\theta} - \bar\kappa(\theta)\, \boldsymbol{\theta},
\end{equation}
where the mean convergence $\bar{\kappa}(\theta)$ (with $\theta = |\boldsymbol{\theta}|$) is
\begin{equation}
    \bar \kappa (\theta) =\frac{1}{D_l^2 \theta^2 \Sigma_{\text{cr}}}\int_0^{D_l \theta} \Sigma(r)\, 2rdr,
\end{equation}
and the critical surface density is $\Sigma_{\text{cr}} = c^2 D_s/(4\pi G D_l D_{ls})$.

\begin{figure}
    \centering
    \includegraphics[width=0.5\linewidth]{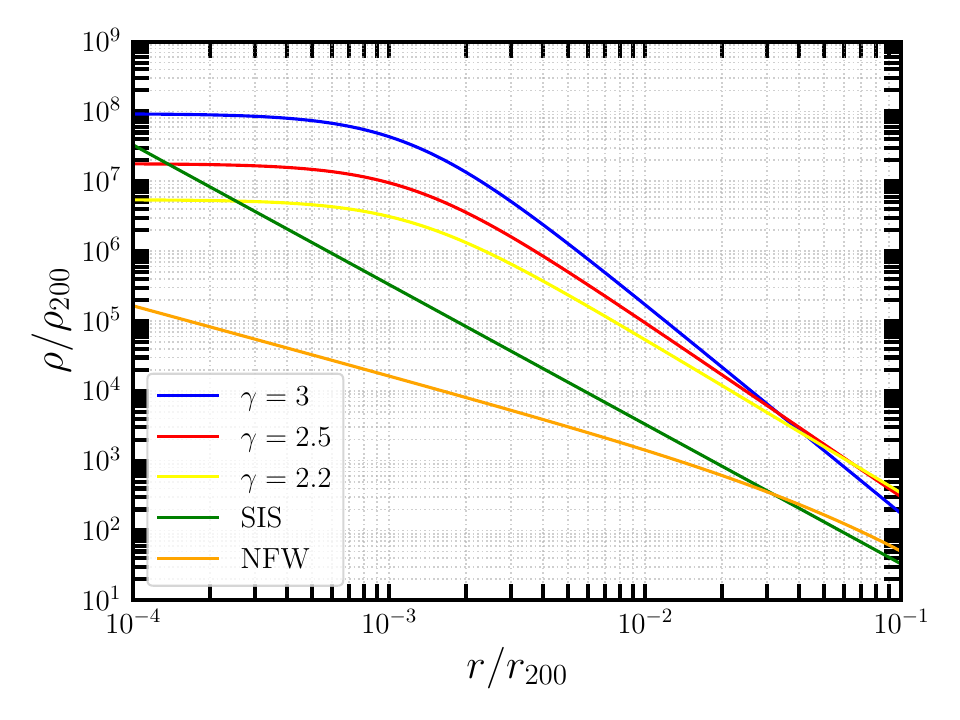}
    \caption{SIS, NFW, and core-collapsed density profiles, with radius normalized to $r_{200}$ and density normalized to $\rho_{200}$, the mean density within $r_{200}$. For a core-collapsed halo with $\gamma=3$, the inner region is significantly denser than other profiles. Note that in core-collapsed cases, we match the mass profile at $r=2r_s$ with that of the NFW profile. }
    \label{fig:profs}
\end{figure}

We consider three density profiles: SIS, NFW (profile~(\ref{eq:nfw})), and the cored power-law profile for core-collapsed halos (profile~(\ref{eq:prof})), shown in Fig.~\ref{fig:profs}. For the SIS profile,
\begin{equation}
    \bar \kappa_{\text{SIS}}(\theta) =  \frac{M^{2/3} (200\pi\rho_c/6)^{2/3}}{\Sigma_{\text{cr}} D_l \theta},
\end{equation}
where $\rho_c$ is the critical density and $M$ is the enclosed mass within $r_{200}$. For the NFW profile,
\begin{equation}
    \bar \kappa_{\text{NFW}} (\theta) = \frac{4 \rho_s r_s^3}{D_l^2\Sigma_{\text{cr}}\theta^2} \left(\ln\frac{D_l \theta}{2 r_s} + F\!\left(\frac{D_l \theta}{r_s}\right)\right),
\end{equation}
where
\begin{equation}
  F(x) = \begin{cases} \sqrt{\dfrac{1}{x^{2}-1}}\tan^{-1}\!\sqrt{x^{2}-1}, & (x > 1), \\ \sqrt{\dfrac{1}{1-x^{2}}}\tanh^{-1}\!\sqrt{1-x^{2}}, & (x < 1), \\ 1, & (x = 1). \end{cases}
\end{equation}
For the core-collapsed profile~(\ref{eq:prof}) with inner slope $\gamma$,
\begin{equation}
    \bar \kappa_{c,\gamma}(\theta)  =
    \begin{cases}
        \dfrac{2\pi^{1/2}\,\Gamma\!\left(\frac{\gamma-1}{2}\right)\rho_0 r_c^3}{\Sigma_{\text{cr}} D_l^2\,\Gamma\!\left(\frac{\gamma}{2}\right)(3-\gamma)\,\theta^2} \left[\left(1+\frac{D_l^2\theta^2}{r_c^2}\right)^{\!\frac{3-\gamma}{2}}-1\right], & (\gamma \neq 3), \\[10pt]
        \dfrac{2\rho_0 r_c^3}{D_l^2\Sigma_{\text{cr}}\,\theta^2}\ln\!\left(1+\frac{D_l^2\theta^2}{r_c^2}\right), & (\gamma = 3).
    \end{cases}
\end{equation}

Within the single-lens regime, we also assess the impact of nearby subhalos on the main lensed (sub)halo. We consider two cases: (A) a subhalo acting as the main lens, perturbed by neighboring subhalos; and (B) a host halo acting as the main lens, perturbed by its own subhalos. Assuming that all halos have SIS profiles for simplicity, the total Fermat potential is
\begin{equation}
\Phi(\boldsymbol{\theta},\boldsymbol{\beta})= \frac{1}{2}|\boldsymbol{\theta} -\boldsymbol{\beta}|^2 -\sum_{j} \psi_j(\boldsymbol{\theta}),\qquad
    \psi_j(\boldsymbol{\theta}) = \kappa_j |\boldsymbol{\theta}-\boldsymbol{\theta}_j|,
\end{equation}
where $\boldsymbol{\theta}_0$ and $\kappa_0$ are the angular position and convergence of the main lens, and $\boldsymbol{\theta}_j$ and $\kappa_j$ ($j = 1, 2, \ldots$) are those of the $j$-th perturbing lens. Treating $\psi_j$ as a perturbation to the main lens $\psi_0$, the gradient for a single perturber ($j=1$) expands as
\begin{equation}
    \nabla_{\boldsymbol{\theta}} \psi(\boldsymbol{\theta})
    = \kappa_0 \frac{\boldsymbol{\theta}}{|\boldsymbol{\theta}|} + \kappa_1\frac{\boldsymbol{\theta}-\boldsymbol{\theta}_1}{|\boldsymbol{\theta}-\boldsymbol{\theta}_1|}
    = \kappa_0 \frac{\boldsymbol{\theta}}{|\boldsymbol{\theta}|} + \frac{\kappa_1}{|\boldsymbol{\theta}_1|^3}\left(|\boldsymbol{\theta}_1|^2\boldsymbol{\theta} - (\boldsymbol{\theta}_1\cdot\boldsymbol{\theta})\boldsymbol{\theta}_1\right) - \kappa_1\frac{\boldsymbol{\theta}_1}{|\boldsymbol{\theta}_1|} + \text{higher-order terms}.
    \label{eq:grad}
\end{equation}
After rotating the $\boldsymbol{\theta}$ plane to eliminate off-diagonal terms, the leading contribution of~(\ref{eq:grad}) reduces to
\begin{equation}
\nabla_{\boldsymbol{\theta}} \psi(\boldsymbol{\theta})
    = \kappa_0 \frac{\boldsymbol{\theta}}{|\boldsymbol{\theta}|} +\begin{pmatrix}
       -\kappa_1/|\boldsymbol{\theta}_1| & 0 \\
       0 & 0
    \end{pmatrix} \boldsymbol{\theta}.
\end{equation}
The caustic size is obtained by solving $\det(\mathrm{Hess}\,\psi(\boldsymbol{\theta})) = 0$ and substituting back into the lens equation~(\ref{eq:le}), yielding
\begin{equation}
\beta_c \approx \kappa_1\kappa_0/|\boldsymbol{\theta}_1| = \beta_0\,\theta_{1E}/|\boldsymbol{\theta}_1|,
\end{equation}
where $\theta_{1E} =\kappa_1$ is the Einstein radius of the perturber. When $|\boldsymbol{\theta}_1| \gg \theta_{1E}$, the caustic is negligible and the lens can be treated as an unperturbed, axially symmetric system. For a subhalo as main lens, and perturbed by the host halo, the ratio $\theta_{1E}/|\boldsymbol{\theta}_1| \sim \theta_{1E}/(r_{200}/D_l) \sim 10^{-6}/(200\,\text{kpc}/1500\,\text{Mpc}) \sim 10^{-2}$ for typical host halos, so the perturbations are negligible in most cases. This argument can be extended to the case of the core-collapsed case with $\gamma=3$ by  replacing $\beta_c \approx \kappa_1\kappa_0/|\boldsymbol{\theta}_1| $ with $\beta_c \approx \kappa_1 \bar{\kappa}_{c,3}(|\boldsymbol{\theta}_1| ) $, which takes a typical value as $\bar{\kappa}_{c,3}(|\boldsymbol{\theta}_1| ) \sim 10^{-4}(r_c/r_{200})/(r_{200}/D_l) \sim 10^{-3}$. The smaller perturbation in the core-collapsed case is mainly due to the density profile dropping faster than SIS away from the center.  For a subhalo perturbed by neighboring subhalos or a host halo perturbed by subhalos, the mean perturbation strength is
\begin{equation}
    \langle \kappa_1/|\boldsymbol{\theta}_1| \rangle \approx \langle\theta_E\rangle/(\langle d \rangle  /D_l) \lesssim 3\times10^{-4} \left(\frac{M}{10^{12}M_\odot}\right)^{0.24},
\end{equation}
  where $\langle d \rangle$ is the typical distance between subhalos. This is negligible for most subhalos with mass below $10^{10}\,M_\odot$. The axially symmetric single-lens approximation is therefore self-consistently justified for the majority of lensing events.

 \section{Observation Efficiency of Lensing Events}

\begin{figure}
    \centering
    \includegraphics[width=0.5\linewidth]{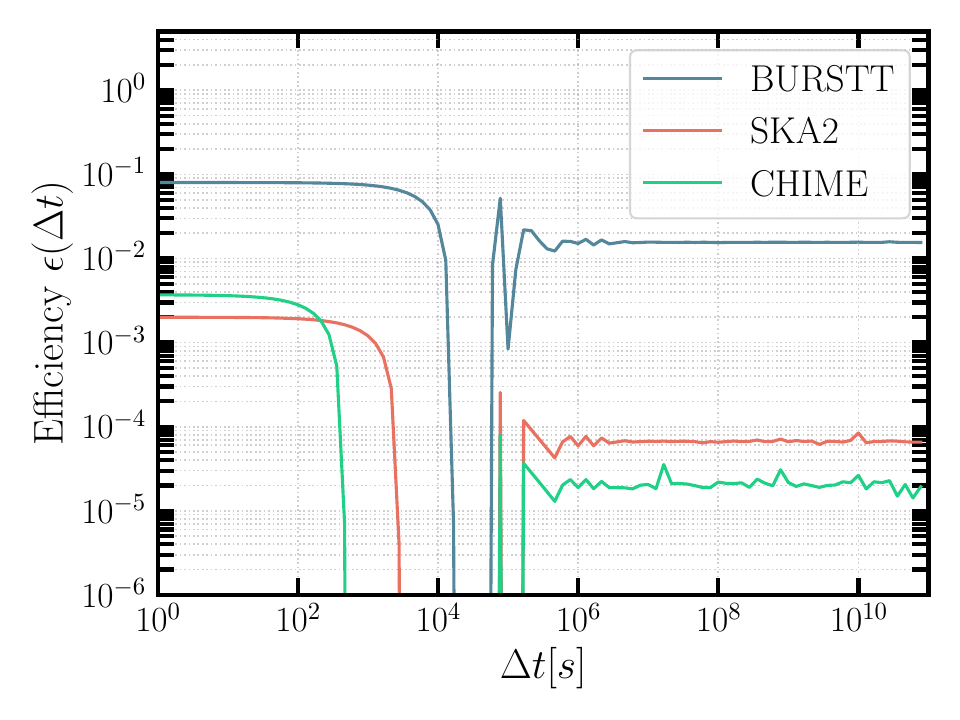}
    \caption{Detection efficiency for lensed images as a function of time delay for BURSTT, SKA2, and CHIME. Efficiency is averaged across the 50 logarithmic time-delay bins by taking 20 points per bin. }
    \label{fig:eff}
\end{figure}

Consider an observatory with field of view $\Delta\Omega$, scan period $T$, and per-scan exposure duration $\Delta T$. Aligning the time origin with the start of the exposure window for a given source, the arrival time $t$ of an observed pulse must satisfy $nT < t < nT + \Delta T$ for some integer $n = 0, 1, 2, \ldots, N$, where $N$ is the total number of scans. For a lensed event, both images must satisfy this condition simultaneously, i.e.,
\begin{equation}
n_1 T < t < n_1 T + \Delta T, \qquad n_2 T < t + \Delta t < n_2 T + \Delta T, \qquad n_1, n_2 = 0, 1, 2, \ldots, N.
\label{eq:time_window}
\end{equation}
This leads to a detection efficiency, defined as the conditional probability that the second image is observed given that the first is observed,
\begin{equation}
    \epsilon(\Delta t) = \left(\frac{\Omega_{\text{FOV}}}{4\pi}\right)\frac{1}{\Delta T} \int_{0}^{\Delta T} dt \sum_{n=0}^N \Theta\!\left[t + \Delta t - nT\right]\Theta\!\left[nT + \Delta T - (t + \Delta t)\right],
    \label{eq:epsilon}
\end{equation}
where $\Theta[x]$ is the Heaviside step function. Writing $\Delta t = kT + \delta t$ with $k = 0, 1, 2, \ldots$ and $0 < \delta t < T$, the product of step functions becomes
\begin{equation}
\Theta\!\left[t + \delta t + (k-n)T\right]\Theta\!\left[(n-k)T + \Delta T - (t+\delta t)\right].
\label{eq:doubletheta}
\end{equation}
For $n = k$, this reduces to $\Theta[\Delta T - \delta t - t]$; for $n - k = 1$, it becomes $\Theta[t + \delta t - T]\,\Theta[T + \Delta T - t - \delta t]$; and it vanishes for all other values of $n$. Substituting into~(\ref{eq:epsilon}) yields
\begin{equation}
    \epsilon(\Delta t) = \epsilon\!\left(\delta t = \Delta t - T\!\left\lfloor\frac{\Delta t}{T}\right\rfloor\right)
    = \begin{cases}
        \left(\dfrac{\Omega_{\text{FOV}}}{4\pi}\right)\dfrac{\Delta T - \delta t}{\Delta T}, & (0 < \delta t < \Delta T), \\[8pt]
        0, & (\Delta T < \delta t < T - \Delta T), \\[8pt]
        \left(\dfrac{\Omega_{\text{FOV}}}{4\pi}\right)\dfrac{\delta t - (T-\Delta T)}{\Delta T}, & (T - \Delta T < \delta t < T),
    \end{cases}
\end{equation}
where $\lfloor\cdot\rfloor$ denotes the floor function. The resulting efficiency curves for the observatories considered in this work are shown in Fig.~\ref{fig:eff}.

\section{Details of Lensing Rate Simulations}

The lens halo distribution is decomposed into host halo and subhalo contributions, 
\begin{equation}
    \frac{d^2n_l}{dz_l dM_l} = \frac{d^2n_{\text{host}}}{dM_{\text{host}} dV}(M_{\text{host}},z_l)\frac{dV}{dz_l}  + \int dM_{\text{host}} \frac{d^2n_{\text{sub}}}{dM_{\text{sub}} dA} \frac{d^2n_{\text{host}}}{dM_{\text{host}} dV}(M_{\text{host}},z_l)\frac{dV}{dz_l} A ,
\end{equation}
where the first term represents the host halo contribution, the second term represents the subhalo contribution, $A = \pi r_{200}^2$ is the projected area within the host halo radius, and $dV/dz_l$ is the comoving volume element at redshift $z_l$.

For host halos, we adopt the velocity dispersion function fitted to a Schechter form from Refs.~\cite{Yue:2021nwt,geng:2021},
\begin{equation}
    \frac{d^2 n_{\text{host}}}{d\ln\sigma_v\, dV} = 6.92\times10^{-3}(1+z)^{-1.18}\,\text{Mpc}^{-3} \left(\frac{\sigma_v}{172.2(1+z)^{0.18}\,\text{km/s}}\right)^{-0.15} \exp\!\left[-\left(\frac{\sigma_v}{172.2(1+z)^{0.18}\,\text{km/s}}\right)^{2.35}\right],
\end{equation}
with the halo mass related to velocity dispersion via $\sigma_v = \left({100\pi G^3 M^2\rho_c}/{3}\right)^{1/6}$.

For subhalos, we adopt the mass function from Refs.~\cite{Gilman:2022ida,Gilman:2019nap},
\begin{equation}
    \frac{d^2n_{\text{sub}}}{dM_{\text{sub}}\,dA} = \frac{\Sigma_{\text{sub}}}{m_0}\left(\frac{M_{\text{sub}}}{m_0}\right)^{\alpha}\left(\frac{M_{\text{host}}}{10^{13}M_\odot}\right)^{k_1}(z+0.5)^{k_2},
\end{equation}
with parameters $m_0 = 10^8\,M_\odot$, $\alpha = -1.85$, $\Sigma_{\text{sub}} = 0.025\,\text{kpc}^{-2}$, $k_1 = 0.55$, and $k_2 = 0.37$, fitted from N-body simulations of CDM. We assume that SIDM follows approximately the same subhalo mass function, since the self-interaction relaxation timescale is much longer than the dynamical timescales that determine the subhalo distribution. We note that tidal interactions mediated by self-interactions between the host halo and its subhalos may lead to corrections to this distribution, which we defer to future work.

The FRB source distribution is
\begin{equation}
    \frac{d^2n_s}{dz_s\,dL_s} = \phi_0\,(1+z_s)^{-1}\frac{dV}{dz_s}\frac{L_s^{\alpha_L-1}}{L_{\text{cutoff}}^{\alpha_L-2}},
\end{equation}
with $\alpha_L = -0.79$, $L_s \in [2\times10^{39},\, 3\times10^{44}]\,\text{erg/s}$, $L_{\text{cutoff}} = 3\times10^{44}\,\text{erg/s}$, and $\phi_0 = 337\,\text{Gpc}^{-3}$. Expanding the exponential in~\eqref{tot} to first order and substituting~\eqref{opt}, the total number of lensing events becomes
\begin{equation}
    N_{\text{tot}} \approx \int dz_s \int dL_s \frac{d^2n_s}{dz_s dL_s} \int 2\pi\beta\,d\beta \int dM_l \int dz_l \frac{d^2n_l}{dz_l dM_l}\,\epsilon(\Delta t)\,
    \Theta\!\left[\min_i\!\left(\mu_i L_s / (4\pi D_s^2(1+z_s)^4)\right) - S_{\text{min}}\right].
    \label{eq:tot_simp}
\end{equation}
We evaluate this integral over $\{M_l, z_l, z_s, L_s, \beta\}$, using the adaptive Monte Carlo package \texttt{vegas}~\cite{Lepage:2020tgj}, which achieves high efficiency with approximately 1\% integration error. The integration ranges are listed in Table~\ref{tab:mcmc}.

\begin{table}[h]
    \centering
    \captionsetup{position=top}
    \caption{Integration ranges for the lensing rate computation in~\eqref{tot_simp}.}
    \begin{tabular}{c|c|c|c|c|c|c}
    \hline
    Parameter & $M_{\text{host}}~[M_\odot]$ & $M_{\text{sub}}~[M_\odot]$ & $z_l$ & $z_s$ & $L_s~[\text{erg/s}]$ & $\beta$ \\
    \hline
    Range & $[10^{10}, 10^{13}]$ & $[10^6, 10^{10}]$ & $[0, z_s]$ & $[0, 3]$ & $[2\times10^{39},\, 3\times10^{44}]$ & $[0, \beta_0]$ \\
    \hline
    \end{tabular}
    \label{tab:mcmc}
\end{table}

To obtain the differential time-delay distribution, we run \texttt{emcee}~\cite{Foreman-Mackey:2012any} with a likelihood proportional to the integrand of~\eqref{tot_simp}, generating samples distributed according to the full multidimensional lensing parameter distribution for both collapsed and non-collapsed cases. For each sample, we compute the time delay between the two brightest FRB images and then reweight the event number by~\eqref{prob} to account for the core-collapse condition and to produce the efficiency-corrected time-delay distribution.

\section{Details of SIDM Parameter Inference}

As a consistency check of the parameter inference pipeline, we generate mock data for the benchmark parameters $\lambda_{\text{sub}} = 0.1$ and $\sigma_{\text{SI}}/m = 10\,\text{cm}^2/\text{g}$, and evaluate the likelihood over the parameter plane as a product of bin-wise Poisson distributions. Core-collapsed halos are assumed to follow profile~(\ref{eq:prof}) with $\gamma = 3$. For subhalos, contributions from non-collapsed subhalos are neglected since the lensing probability of NFW or cored halos is negligibly small. For host halos, non-collapsed halos are assumed to follow SIS profiles. The likelihood is
\begin{equation}
    \mathcal{L}(\lambda_{\text{sub}}, \sigma_{\text{SI}}/m) = \prod_i \frac{n_i(\lambda_{\text{sub}}, \sigma_{\text{SI}}/m)^{N_i}\,e^{-n_i(\lambda_{\text{sub}}, \sigma_{\text{SI}}/m)}}{N_i!},
\end{equation}
where $N_i$ is the observed number of lensed events in the $i$-th bin from the mock data and $n_i(\lambda_{\text{sub}}, \sigma_{\text{SI}}/m)$ is the corresponding expected number.

\begin{figure*}
    \centering
    \includegraphics[width=0.32\linewidth]{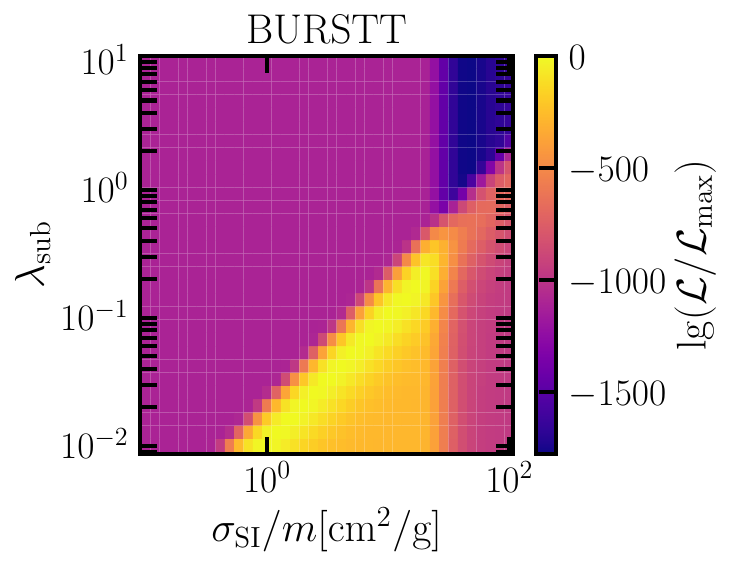}
    \includegraphics[width=0.32\linewidth]{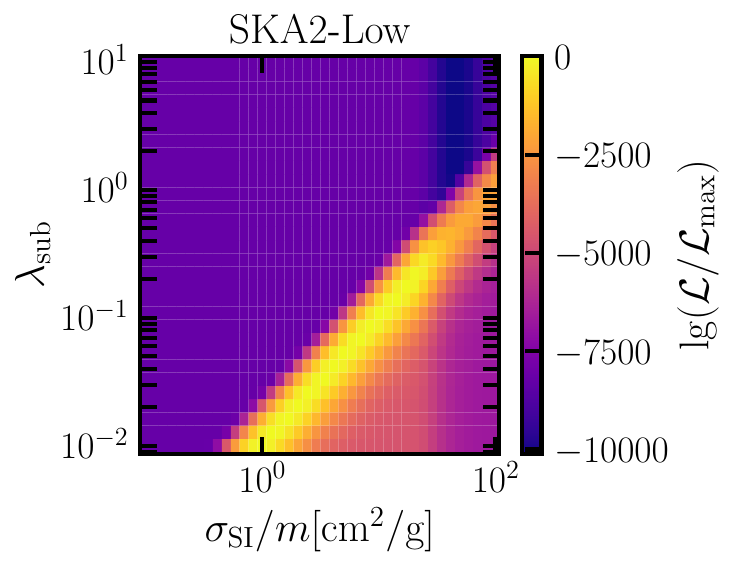}
    \includegraphics[width=0.32\linewidth]{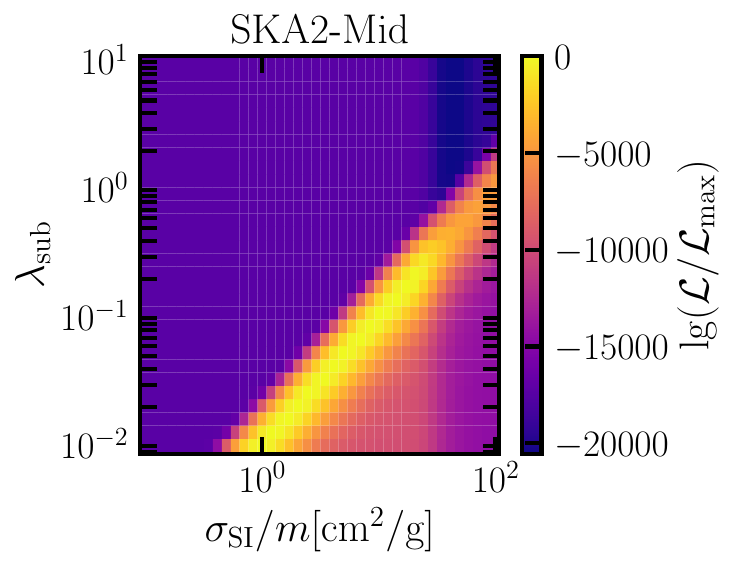}
    \caption{Relative log-likelihood in the $\lambda_{\text{sub}}$--$\sigma_{\text{SI}}/m$ plane derived from mock time-delay data for BURSTT, SKA2-Low, and SKA2-Mid.}
    \label{fig:total}
\end{figure*}

Figure~\ref{fig:total} shows the resulting relative log-likelihood for BURSTT, SKA2-Low, and SKA2-Mid. As expected, the constraining power is primarily determined by the total number of observed events, with SKA2-Mid achieving the best sensitivity due to its lowest flux threshold $S_{\text{min}}$. The likelihood peaks near the injected benchmark values ($\lambda_{\text{sub}} = 0.1$, $\sigma_{\text{SI}}/m = 10\,\text{cm}^2/\text{g}$), and a parameter degeneracy appears along the direction $\sigma_{\text{SI}}/m \sim 40\lambda_{\text{sub}}\,\text{cm}^2/\text{g}$, consistent with the exclusion contours in the main text (see~\figref{proj}) and confirming the self-consistency of the inference procedure.

\end{document}